\documentclass[onecolumn,showpacs,preprintnumbers,amsmath,amssymb]{revtex4}
\usepackage{graphicx}
\usepackage{dcolumn}
\usepackage{bm}
\begin{document}

\newcommand{\kvec}{\mbox{{\scriptsize {\bf k}}}}
\newcommand{\lvec}{\mbox{{\scriptsize {\bf l}}}}
\newcommand{\qvec}{\mbox{{\scriptsize {\bf q}}}}
%
\def\eq#1{(\ref{#1})}
\def\fig#1{\hspace{1mm}\ref{#1}}
\def\tab#1{\hspace{1mm}\ref{#1}}

\title{The toy model for the high-$\rm{T_{C}}$ superconductivity}

\author{R. Szcz{\c{e}}{\`s}niak}
\email{szczesni@wip.pcz.pl}
\affiliation{Institute of Physics, Cz{\c{e}}stochowa University of Technology, Al. Armii Krajowej 19, 42-200 Cz{\c{e}}stochowa, Poland}
\date{\today}
\begin{abstract}
The simple microscopic model for the high-$T_{C}$ superconductors based on the electron-phonon (EPH) and electron-electron-phonon (EEPH) interactions has been presented.
On the {\it fold} mean-field level, it has been shown, that the obtained model supplements the predictions based on the BCS van Hove scenario. In particular:
(i) For strong EEPH coupling and $T<T_{C}$ the energy gap ($\Delta_{tot}$) is very weak temperature dependent; up to the critical temperature $\Delta_{tot}$ extends into the anomalous normal state to the Nernst temperature. 
(ii) The model explains well the experimental dependence of the ratio $R_{1}\equiv 2\Delta_{tot}^{\left(0\right)}/k_{B}T_{C}$ on doping for the reported superconductors in the terms of the few fundamental parameters.
\end{abstract}
\maketitle
\section{\label{sec:1}INTRODUCTION}

In the study, we present the simple microscopic theory of the high-$T_{C}$ superconductivity \cite{Bednorz}. 
The organization of the paper is as follows: 

In Section II we call for the pairing mechanism. First of all, we discuss the main experimental and theoretical results. Next, on the basis of the presented analysis, we give the postulates, which determine the microscopic model for the high-$T_{C}$ superconductors in the second quantization form.

In Section III, by using the unitary transformation, we deduce from the initial Hamiltonian the simple mean-field model. In the framework of the toy model (only the {\it s}-wave state), we discuss the properties of the energy gap in the superconducting and Nernst region; the numerical predictions are supplemented by the analytical approach.  Next, for selected high-$T_{C}$ superconductors, we calculate the fundamental parameters of the model and compare the obtained theoretical results with the experimental data. 

Finally, in Section IV we summarize the results.

The main body of the paper is supplemented by the Appendixes. The formally exact expression for the self-energy matrix is obtained in Appendix A. In Appendix B we present in detail the {\it fold} mean-field approximation. The van Hove and generalized mean-field thermodynamic potentials are calculated in Appendix C. The lists of the experimental values of the thermodynamic parameters for the selected high-$T_{C}$ superconductors are collected in Appendix D.

\section{THE PAIRING MECHANISM}
%
\begin{figure}[t]%
\includegraphics*[scale=1]{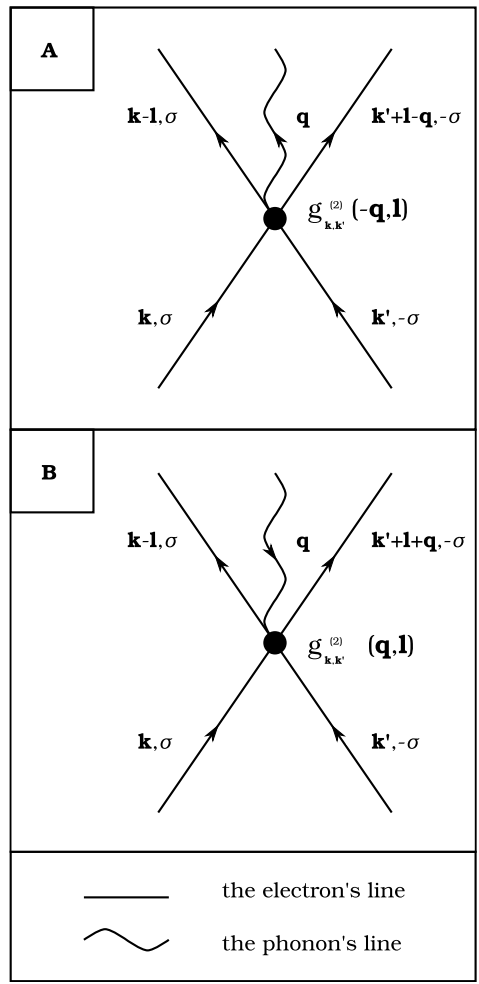}
\caption{The graphical representation of the electron-electron-phonon coupling. The electron states are represented by the straight lines, while the phonon state by the curly line. The vertex is denoted by the black dot and its strength is given by $g^{\left(2\right)}_{\kvec,\kvec^{'}}\left({\bf q},{\bf l}\right)$. In Fig.\fig{f1} (A) the electrons emit the phonon during the scattering; in Fig.\fig{f1} (B) the electrons absorb the phonon.} 
\label{f1} 
\end{figure}
%

The real cuprates are three-dimensional. However, their physical properties are strongly
anisotropic. On the basis of very small coherence length in the {\it c} direction (smaller than the interplane distance) one can suppose that the ${\rm{CuO_{2}}}$  electrons play the special role in the physics of the high-$T_{C}$ superconductors \cite{Dagotto}. Unfortunately, the pairing mechanism for the planar problem remains highly controversial and many different hypotheses are suggested. In the literature two fundamental directions in search for the pairing mechanism have been crystallized. The first approach is based on the single-band Hubbard model, its extensions or related models {\it e.g.} $t-J$ model \cite{ Hubbard}-\cite{Spalek}; the second approach emphasizes the relevance of the electron-phonon interaction \cite{Kulic}.

Why is the Hubbard model so much studied? First of all, some analysis suggest that the one-band Hubbard model reproduces well the spectra of the more complicated three-band Hamiltonian for electrons in the copper oxide planes (the Emery model) \cite{Dagotto}, \cite{Emery1}. For example, by using the finite cluster method, Hybertsen {\it et al.} have shown that the one-band Hubbard model with the small next-nearest-neighbor integral $t^{'}$ should have the following parameters: $t=430$ meV, $t^{'}=-70$ meV and $U_{H}=5.4$ eV \cite{Hybertsen}. Secondly, for the half-filled electron band and large on-site Coulomb interaction, the Hubbard model reduces to the Heisenbeg model, which describes well the spin dynamics of the underdoped high-$T_{C}$ superconductors \cite{Dagotto}. On the basis of the quoted facts some authors suppose, that the strong electronic correlations modeled by the Hubbard model can alone induce the superconducting state in the cuprates. Unfortunately, the studies carried out by several groups have shown that the Hubbard model gives no obvious evidence for superconductivity with the large critical temperature \cite{Imada}. On the other hand, there is the strong tendency for superconductivity in the attractive Hubbard model for the same value of the on-site Coulomb interaction. Finally, we noticed that probably also the three-band Hubbard model and the $t-J$ model do not superconduct at temperatures characteristic for the cuprates \cite{Dagotto}, \cite{Pryadko}.

The relevance of phonons to the pairing mechanism in the high-$T_{C}$ superconductivity also constitutes a complicated problem. On the one hand there exist many experimental observations which have been taken as evidence for the electron-phonon interaction in the cuprates. For example: the strong isotope effects on $T_{C}$  in the underdoped superconductors \cite{Franck}, the phonon renormalization in the Raman measurements \cite{Kulic2}, the phonon-related features of  $I-V$ characteristics obtained by using the tunnelling experiments \cite{Vedeneev} and the dependence of  the penetration depth on the substitution ${\rm O^{16}}$  by  ${\rm O^{18}}$ \cite{Hofer}. Especially important results come from the ARPES measurements which give the evidence on the low-energy kink in the quasiparticle spectrum around the phonon energy both for the nodal and antinodal points \cite{Damascelli}, \cite{Cuk}; also the ARPES isotope effect in ${\rm Re}\left(\Sigma\right)$ has been observed \cite{Gweon}. On the other hand the first principles calculations support the view that the conventional electron-phonon coupling is small \cite{Heid}.  For example, Bohnen {\it et al.} have predicted that the electron-phonon coupling constant for ${\rm YBa_{2}Cu_{3}O_{7-y}}$ is equal to $0.27$; so the strong Hubbard correlations should completely suppress the phonon-mediated superconductivity \cite{Bohnen}.

After summarizing the mentioned experimental and theoretical results, one can conclude that: (i) the cuprates belong to the strongly correlated systems but probably these correlations in the superconductivity domain are beyond the Hubbard or related approaches, since in these models the pairing correlations are too small, (ii) in the cuprates the conventional electron-phonon interaction is small but according to the experimental data one can suppose that the phonons play the important role in the pairing mechanism.  

In order to solve the problem of high temperature superconductivity we present and examine the following scenario:\\ 
(i) {\it In the superconductivity domain of the cuprates the fundamental role is played by the electrons on the $CuO_{2}$ planes}.\\
(ii) {\it In the cuprates exists the conventional electron-phonon interaction, which has not to be strong}.\\
(iii) {\it In the cuprates exist strong electronic correlations, but the electron-electron scattering in the superconductivity domain is inseparably connected with the absorption or emission of vibrational quanta.}

In the simplest case the first and second postulate coincides with the phonon-van-Hove-scenario for high-$T_{C}$ superconductors \cite{VanHove}, \cite{Markiewicz}. The third postulate states that the effective electronic correlations in the superconductivity domain are connected with the three-body process: the electron-electron-phonon interaction. In Fig. \fig{f1} we show in detail the diagrammatic representation of this interaction. We notice that the EEPH coupling has a significant property which distinguish it from the Hubbard interaction; it does not destroy the classical phonon-mediated pairing correlations. Additionally, one should pay attention to the fact, that the first postulate has also the essential significance for the third postulate. Namely, for the two-dimensional case (the van Hove singularity at the Fermi level), the EEPH coupling has significantly strong influence on the physical properties of the system (see next section).  

Below, we consider the Hamiltonian that describes the postulated coupling of the correlated electrons to phonons in the second quantization form:
\begin{equation}
\label{r1(2.0)}
H\equiv H^{\left(0\right)}+H^{\left(1\right)}+H^{\left(2\right)}.
\end{equation}
The first term represents the non-interacting electrons and phonons:
\begin{equation}
\label{r2(2.0)}
H^{\left(0\right)}\equiv\sum_{\kvec\sigma }\overline\varepsilon _{\kvec}c_{\kvec\sigma
}^{\dagger}c_{\kvec\sigma }+\sum_{\qvec}\omega _{\qvec}b_{\qvec}^{\dagger}b_{\qvec},
\end{equation}
where $\overline\varepsilon _{\kvec}\equiv \varepsilon _{\kvec}-\mu $; $\varepsilon _{\kvec}$ and $\mu$ denotes the electron band energy and the chemical potential respectively. For the two-dimensional square lattice and the nearest-neighbor hopping integral $t$, we have: 
$\varepsilon _{\kvec}=-t\gamma\left({\bf k}\right)$, where
$\gamma\left({\bf k}\right)\equiv2\left[\cos\left(k_{x}\right)+\cos\left(k_{y}\right)\right]$.
The symbol $\omega_{\qvec}$ stands for the energy of phonons.
The interaction terms are given by:
\begin{equation}
\label{r3(2.0)}
H^{\left(1\right)}\equiv\sum_{\kvec\qvec\sigma }g^{\left(1\right)}_{\kvec}\left({\bf q}\right)
c_{\kvec+\qvec\sigma}^{\dagger}c_{\kvec\sigma}\phi_{\qvec},
\end{equation}
and
\begin{equation}
\label{r4(2.0)}
H^{\left(2\right)}\equiv\sum_{\kvec\kvec^{'}\qvec\lvec\sigma}
g^{\left(2\right)}_{\kvec,\kvec^{'}}\left({\bf q},{\bf l}\right)
c_{\kvec-\lvec\sigma }^{\dagger}c_{\kvec\sigma}
c_{\kvec^{'}+\lvec+\qvec-\sigma}^{\dagger}c_{\kvec^{'}-\sigma}\phi_{\qvec},
\end{equation}
where $\phi_{\qvec}\equiv b_{-\qvec}^{\dagger}+b_{\qvec}$.
The matrix elements $g^{\left(1\right)}_{\kvec}\left({\bf q}\right)$ 
describe the electron-phonon interaction \cite{Frohlich} and the symbol 
$g^{\left(2\right)}_{\kvec,\kvec^{'}}\left({\bf q},{\bf l}\right)$ 
determines the strength of the electron-electron-phonon coupling. Since the Hamiltonian is the hermitian operator, we have: 
$g^{\star\left(1\right)}_{\kvec}\left({\bf q}\right)=
g^{\left(1\right)}_{\kvec+\qvec}\left({\bf -q}\right)$ and $g^{\star\left(2\right)}_{\kvec,\kvec^{'}}\left({\bf q},{\bf l}\right)=g^{\left(2\right)}_{\kvec^{'}+\lvec+\qvec,\kvec-\lvec}\left({\bf -q},{\bf l+q}\right)$.

\section{THE FOLD MEAN-FIELD THEORY}

From the mathematical point of view, the Hamiltonian \eq{r1(2.0)} belongs to the class of so-called {\it dynamic Hubbard models}. These Hamiltonians describe modulation of the Hubbard on-site repulsion by the boson degree of freedom and were  studied in detail by J.E. Hirch, F. Marsiglio and R. Teshima in the papers \cite{Hirsch}-\cite{Marsiglio}.

In the simplest case, the dynamic Hubbard Hamiltonian describes the Holstein-like electron-boson interaction, where the coupling constant increases with the electron occupation. This model, however interesting, is inappropriate in the description of the high-$T_{C}$ superconductivity.
In the more elaborate scheme, the properties of the dynamic Hubbard model, can be studied by using the generalized Lang-Firsov transformation: $H^{'}=e^{S}He^{-S}$, where the generator $S$ has the form: $S\equiv\sum_{j}g\left(b_{j}-b^{\dagger}_{j}\right)\left(n_{j\uparrow}+n_{j\downarrow}- n_{j\uparrow}n_{j\downarrow}\right)$ and $n_{j\sigma}$ is the number operator for the site $j$ and the spin $\sigma$. Next, the obtained effective Hamiltonian should be analyzed in the framework of the Eliashberg-like approximation \cite{Marsiglio}-\cite{Eliashberg}; the central result is that the critical temperature increases with retardation.

We notice that the form of the Hamiltonan \eq{r1(2.0)} suggests using the Eliashberg approach in order to analyse the physical properties of the considered system \cite{Eliashberg}. This method is appropriate, because one can retain simultaneously the electron and phonon degrees of freedom. Additionally, in the framework of the Eliashberg formalism, the formally exact expression for the self-energy matrix is easy to determine. The analysis of the above issue is presented in Appendix A, where we have shown that: (i) in the case of the absence of the lattice distortion, the first-order contribution to the self-energy is equal to zero, (ii) the electron-phonon coupling is directly increased by the electron-electron-phonon interaction, (iii) the form of the second-order contributions to the matrix self-energy is very complicated. Unfortunately, this fact hampers considerably the detailed calculations on the Eliashberg level. 

In the case of the presented paper, we have eliminated the phonon operators only from the EEFH term. The obtained Hamiltonian is next added to the BCS operator \cite{Bardeen}. As discussed in the following subsections, presented approximate description represents a generalization of the BCS van Hove scenario \cite{Markiewicz}.

\subsection{The unitary transformation}

%
\begin{figure}[t]%
\includegraphics*[scale=1]{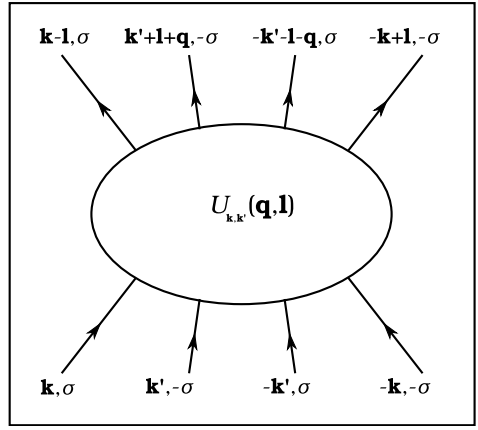}
\caption{The four-body scattering event contributing to the interaction part of the Hamiltonian \eq{r6(3.1)}. The oval connecting the electron's lines is an illustration of 
$U_{\kvec,\kvec^{'}}\left({\bf q},{\bf l}\right)$ which denotes the effective potential.}
\label{f2} 
\end{figure}
%

In order to eliminate the phonon degrees of freedom in the Hamiltonian \eq{r4(2.0)} we use the Fr{\"o}hlich-type unitary transformation: 
\begin{equation}
\label{r1(3.1)}
H^{'}\equiv e^{-iS}\left(H^{\left(0\right)}+H^{\left(2\right)}\right)e^{iS},
\end{equation}
where the operator $S$ denotes the generator:
\begin{equation}
\label{r2(3.1)}
S\equiv\sum_{\qvec}S_{\qvec}.
\end{equation}
In our case: $S_{\qvec}\equiv\gamma_{\qvec}b_{\qvec}+\gamma^{\dagger}_{\qvec}b^{\dagger}_{\qvec}$, where
\begin{equation}
\label{r3(3.1)}
\gamma_{\qvec}\equiv\sum_{\kvec\kvec^{'}\lvec\sigma }
\Phi\left({\bf k},{\bf k^{'}},{\bf l},{\bf q}\right)
c_{\kvec-\lvec\sigma }^{\dagger}c_{\kvec\sigma}
c_{\kvec^{'}+\lvec+\qvec-\sigma }^{\dagger}c_{\kvec^{'}-\sigma}.
\end{equation}
The expression \eq{r1(3.1)} can be rewritten in the approximate form:
\begin{eqnarray}
\label{r4(3.1)}
H^{'}&\simeq& H^{\left(0\right)}
+\left(i[H^{\left(0\right)},S]_{-}+H^{\left(2\right)}\right)\\ \nonumber
&+&i\left[\left(\frac{i}{2}\left[H^{\left(0\right)},S\right]_{-}
+H^{\left(2\right)}\right),S\right]_{-},
\end{eqnarray}
where the square brackets $\left[,\right]_{-}$ denote the commutator.
To eliminate the second term in Eq. \eq{r4(3.1)} we assume that the generator fulfills the relation: 
$i[H^{\left(0\right)},S]_{-}+H^{\left(2\right)}=0$, hence: 
\begin{equation}
\label{r5(3.1)}
\Phi\left({\bf k},{\bf k^{'}},{\bf l},{\bf q}\right)=\frac{ig^{\left(2\right)}_{\kvec,\kvec^{'}}\left({\bf q},{\bf l}\right)}
{\varepsilon_{\kvec-\lvec}-\varepsilon_{\kvec}+\varepsilon_{\kvec^{'}
+\lvec+\qvec}-\varepsilon_{\kvec^{'}}-\omega_{\qvec}}.
\end{equation}
Next, we can reduce the Hamiltonian $H^{'}$ to the following expression:
\begin{widetext}
\begin{equation}
\label{r6(3.1)}
H_{eff}\equiv <0_{{\rm ph}}|H^{'}|0_{{\rm ph}}>\simeq\sum_{\kvec\sigma }
\overline\varepsilon _{\kvec}c_{\kvec\sigma}^{\dagger}c_{\kvec\sigma }
+\sum_{\kvec\kvec^{'}\qvec\lvec\sigma}
U_{\kvec,\kvec^{'}}\left({\bf q},{\bf l}\right)
c_{\kvec-\lvec\sigma }^{\dagger}c_{\kvec\sigma}
c_{\kvec^{'}+\lvec+\qvec-\sigma }^{\dagger}c_{\kvec^{'}-\sigma}
c_{-\kvec^{'}-\lvec-\qvec\sigma }^{\dagger}c_{-\kvec^{'}\sigma}
c_{-\kvec+\lvec-\sigma }^{\dagger}c_{-\kvec-\sigma},
\end{equation}
\end{widetext}
where the phonon vacuum state is given by $|O_{\rm{ph}}>$.
The pairing potential $U_{\kvec,\kvec^{'}}\left({\bf q},{\bf l}\right)$ is of the form:
\begin{equation}
\label{r7(3.1)}
U_{\kvec,\kvec^{'}}\left({\bf q},{\bf l}\right)\simeq
\frac{\omega_{0}|g^{\left(2\right)}|^{2}}
{\left(\varepsilon_{\kvec}-\varepsilon_{\kvec-\lvec}+\varepsilon_{\kvec^{'}}-
\varepsilon_{\kvec^{'}+\lvec+\qvec}\right)^{2}-\omega^{2}_{0}}.
\end{equation}
We assume additionally that: $g^{\left(2\right)}_{\kvec,\kvec^{'}}\left({\bf q},{\bf l}\right)\simeq g^{\left(2\right)}$ and $\omega_{\qvec}\simeq\omega_{0}$; the symbol $\omega_{0}$ denotes the characteristic phonon frequency.
On the basis of the expression \eq{r6(3.1)} we conclude that the EEPH interaction can be replaced by the effective four electron-electron (4EE) scattering event; the diagrammatic representation of this interaction is shown in Fig.\fig{f2}. 
From the Eq. \eq{r7(3.1)} it is clear that the effective potential is attractive if:
$|\varepsilon_{\kvec}-\varepsilon_{\kvec-\lvec}+\varepsilon_{\kvec^{'}}-\varepsilon_{\kvec^{'}+\lvec+\qvec}|<\omega_{0}$.  
In the subsection, we consider the simplest case, when the attractive part of the 4EE potential can be written as: 
\begin{equation}
\label{r8(3.1)}
U^{\left(<\right)}_{\kvec,\kvec^{'}}\left({\bf q},{\bf l}\right)\rightarrow -\frac{U}{24N^{3}},
\end{equation}
for $|\varepsilon_{\kvec}|<\omega_{0}$. We have used the factor $\frac{1}{24}$, because the potential energy term represents the interaction between every four of particles counted once.

By using the {\it fold} mean-field (MF) approximation the Hamiltonian \eq{r6(3.1)} takes the form (see also Appendix B):
\begin{eqnarray}
\label{r9(3.1)}
H_{eff}^{MF}&=&
\sum_{\kvec\sigma }
\overline\varepsilon _{\kvec}c_{\kvec\sigma}^{\dagger}c_{\kvec\sigma }\\ \nonumber
&-&\frac{U}{12}\sum^{\omega_{0}}_{\kvec\sigma}
|\Delta_{\sigma}|^{2}
\left(
\Delta_{\sigma}c^{\dagger}_{\kvec-\sigma}c^{\dagger}_{-\kvec\sigma}
+
\Delta^{\star}_{\sigma}c_{-\kvec\sigma}c_{\kvec-\sigma}
\right).
\end{eqnarray}
The symbol $\sum^{\omega_{0}}_{\kvec}$ denotes the sum over the states when the 4EE potential is attractive; $\Delta_{\sigma}\equiv\frac{1}{N}\sum^{\omega_{0}}_{\kvec}
\left<c_{-\kvec\sigma}c_{\kvec-\sigma}\right>$.

\subsection{The toy model}

The thermodynamic parameters of the high-$T_{C}$ superconductors can have essentially different properties in comparison with the low-$T_{C}$  materials. From this reason, the analysis of the results obtained in the framework of the simplest approach is important, because these predictions facilitate the interpretation of the fundamental experimental data. 

Taking into account the operator \eq{r9(3.1)} we can write the total Hamiltonian in the form:
\begin{eqnarray}
\label{r1(3.2)}
& &H^{MF}\equiv\sum_{\kvec\sigma}\varepsilon _{\kvec}c_{\kvec\sigma}^{\dagger}c_{\kvec\sigma}\\ \nonumber
&-&\left(V+\frac{U}{6}|\Delta|^{2}\right)\sum^{\omega_{0}}_{\kvec}
\left(\Delta c^{\dagger}_{\kvec\uparrow}c^{\dagger}_{-\kvec\downarrow}+
\Delta^{\star}c_{-\kvec\downarrow}c_{\kvec\uparrow}\right),
\end{eqnarray}
where we have omitted the chemical potential and $\Delta\equiv\Delta_{\downarrow}$. In Eq. \eq{r1(3.2)}, the symbol $V$ represents the BCS pairing potential obtained from the Hamiltonian \eq{r3(2.0)}.

Now, we establish the energy scales in the presented model.  The nearest-neighbor hopping integral $t$ is of the order of ($200-400$) meV 
\cite{Xu}-\cite{Lin}. We notice that in the numerical model calculations we take $t$ as an energy unit. 
From the {\it ab initio} calculations arises the fact that $V\in\left(0,2t\right)$ \cite{Bohnen}.
In order to determine qualitatively the possible values of $U$, we note, that the simple BCS pairing potential is obtained from the expression:
\begin{equation}
\label{r2(3.2)}
\frac{\omega_{0}|g^{\left(1\right)}|^{2}}{\left(\varepsilon_{\kvec}-\varepsilon_{\kvec+\qvec}\right)^{2}-\omega^{2}_{0}}\rightarrow -\frac{V}{2N},
\end{equation}
where the characteristic phonon frequency $\omega_{0}$ is of the order of Debye frequency ($\omega_{D}\sim 0.3t$); the electron-phonon coefficient $g^{\left(1\right)}$ has nearly the same value \cite{Hauge}. 
Next, we assume that the largest energy in the high-$T_{C}$ superconductors, of order $\left(5-10\right)$ eV, is the electron-electron-phonon potential; in the other words $g^{\left(2\right)}$ is comparable with the Coulomb repulsion in the one-band Hubbard model. Hence, $g^{\left(2\right)}/ g^{\left(1\right)}\sim 10^{2}$. 
Then, by using Eqs. \eq{r7(3.1)}, \eq{r8(3.1)} and \eq{r2(3.2)} we can calculate the ratio $U/V$. The result shows that $U/V$ is proportional to  $10\left(g^{\left(2\right)}/g^{\left(1\right)}\right)^{2}$. Due to this reason $U/V$ can be considerably larger than $g^{\left(2\right)}/g^{\left(1\right)}$. 

From the mathematical point of view, the value of $U$ is not as much important as the value of the mean-field potential ($U_{MF}\equiv \frac{U}{6}|\Delta|^{2}$). It is easy to see that, in contrast to the BCS pairing potential, $U_{MF}$ is $\Delta$-dependent (the fundamental feature of the presented model). So, the potential $U_{MF}$ depends on the temperature, $V$ and $U$. If we set $V$ and $U$ the mean-field potential reaches the maximum value for $T=0$ K: 
$U^{\left(0\right)}_{MF}\equiv\left[U_{MF}\right]_{T=0}=\frac{U}{6}|\Delta^{\left(0\right)}|^{2}$, where the symbol $|\Delta^{\left(0\right)}|$ denotes the amplitude of the anomalous thermal average for $T=0$ K. In the case of the analyzed superconductors, we have $\left[U^{\left(0\right)}_{MF}\right]_{\rm max}\simeq 3t$ (see also subsection C).

By using the Hamiltonian \eq{r1(3.2)} we calculate the anomalous Green function:
\begin{equation}
\label{r3(3.2)}
\left<\left<c_{\kvec\uparrow}|c_{-\kvec\downarrow}\right>\right>=-
\frac{\left(V+\frac{U}{6}|\Delta|^{2}\right)\Delta}{\omega^{2}-E^{2}_{\kvec}},
\end{equation}
where $E_{\kvec}\equiv\sqrt{\varepsilon_{\kvec}^{2}+\left(V+\frac{U}{6}|\Delta|^{2}\right)^{2}|\Delta|^{2}}$. The obtained propagator is more complicated than the BCS Green function; peculiarly, we draw the readers' attention to the very intricate structure of the energy gap. 
The fundamental equation can be found in the form:
\begin{equation}
\label{r4(3.2)}
1=\left(V+\frac{U}{6}|\Delta|^{2}\right)\frac{1}{N}\sum^{\omega_{0}}_{\kvec}
\frac{1}{2E_{\kvec}}\tanh\frac{\beta E_{\kvec}}{2}.
\end{equation}
In order to calculate the thermodynamic properties we transform the momentum summation over an energy integration in Eq.\eq{r4(3.2)}. We notice that, in the case of three-dimensional system, where the electron density of states near the Fermi energy is constant, the mean-field 4EE interaction can be neglected because the value of $U^{\left(0\right)}_{MF}$ is very small. The situation changes dramatically for the two-dimensional system, where $U^{\left(0\right)}_{MF}$ can be even greater than $V$. Then:
\begin{equation}
\label{r5(3.2)}
1=V_{tot}\int^{\omega_{0}}_{0}
d\varepsilon\rho\left(\varepsilon\right)
\frac{\tanh\left(\frac{\beta}{2}E\right)}{E},
\end{equation}
where: $E\equiv\sqrt{\varepsilon^{2}+\Delta^{2}_{tot}}$, $\Delta_{tot}\equiv V_{tot}|\Delta|$ and $V_{tot}\equiv V+\frac{U}{6}|\Delta|^{2}$. In the case of the square lattice the density of states is given by \cite{Szczesniak1}-\cite{Mamedov}:
\begin{equation}
\label{r6(3.2)}
\rho\left(\varepsilon\right)=b_{1}\ln|\frac{\varepsilon}{b_{2}}|,
\end{equation}
where $b_{1}=-0.04687t^{-1}$ and $b_{2}=21.17796t$.

%
\begin{figure}[t]%
\includegraphics*[scale=0.15]{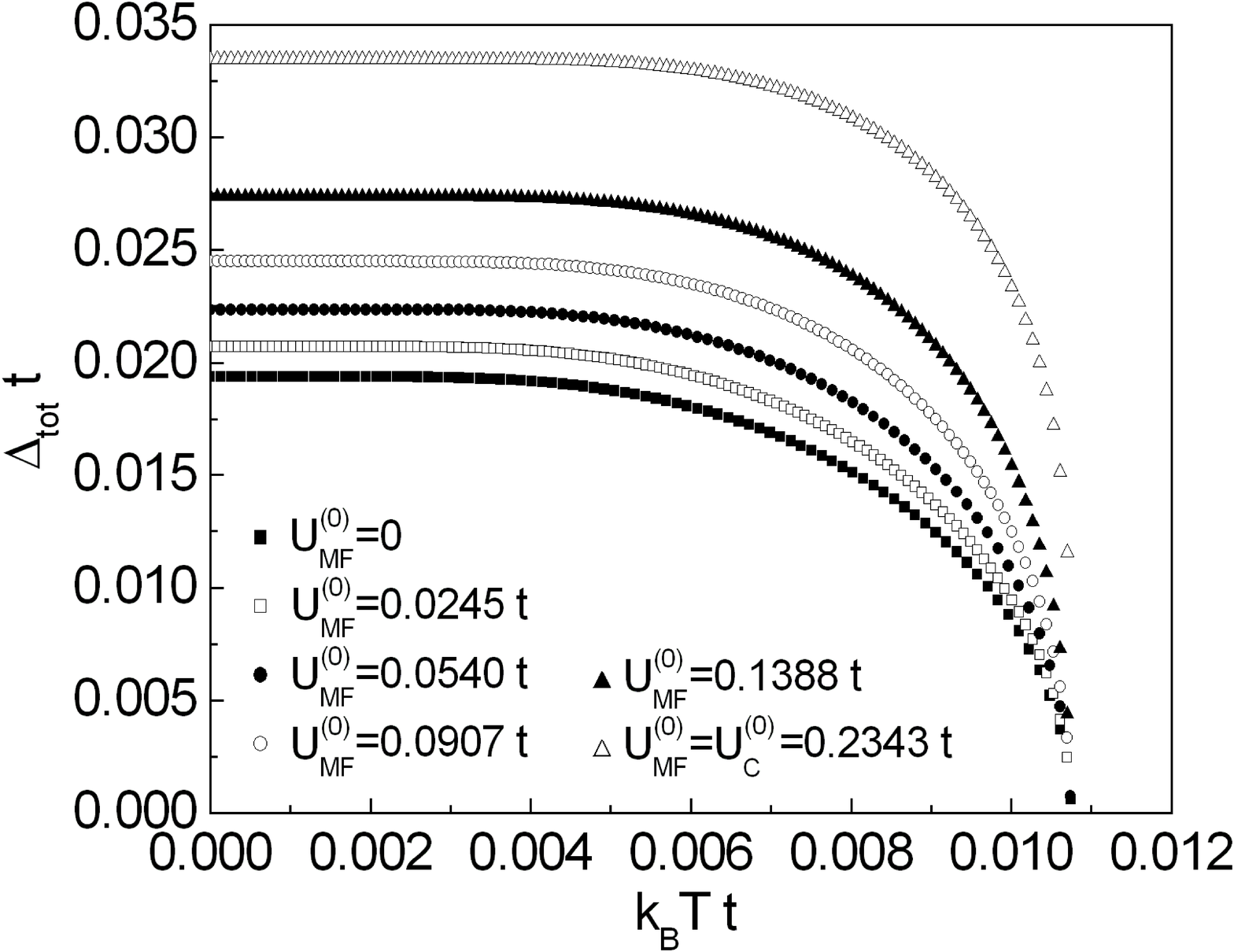}
\caption{The dependence of $\Delta_{tot}$ on the temperature for $U^{\left(0\right)}_{MF}\leq U^{\left(0\right)}_{C}=0.2343t$. We assume $V=1t$ and $\omega_{0}=0.3t$.}
\label{f3} 
\end{figure}
%
\begin{figure}[t]%
\includegraphics*[scale=0.15]{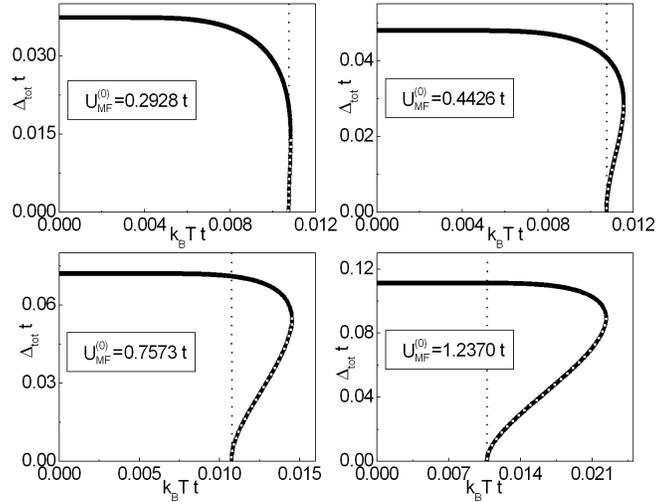}
\caption{The dependence of $\Delta_{tot}$ on the temperature for $U^{\left(0\right)}_{MF}>U^{\left(0\right)}_{C}$. We assume $V=1t$ and $\omega_{0}=0.3t$. The vertical line indicates a position of the critical temperature. For  $T>T_{C}$ the solid line represents the higher branch, whereas the perforated line corresponds to the lower branch.}
\label{f4} 
\end{figure}
%
 
In Figs. \fig{f3} and \fig{f4} we show the numerical solutions of Eq. \eq{r5(3.2)} for increasing values of $U^{\left(0\right)}_{MF}$. Analysis of the presented results allows one to state that only for $U^{\left(0\right)}_{MF}\leq U^{\left(0\right)}_{C}$ (where $U^{\left(0\right)}_{C}$ is some characteristic value) the gap equation has one solution. Above $U^{\left(0\right)}_{C}$  at $T_{C}$ open the two new branches of  the energy gap. 

We notice that in the framework of the obtained mean-field model the 4EE interaction does not influence on the value of $T_{C}$; so, the critical temperature can be calculated by using the expression \cite{Szczesniak1}:
\begin{equation}
\label{r7(3.2)}
k_{B}T_{C}=ab_{2}e^{-\frac{1}{\lambda_{1}}},
\end{equation}
where
\begin{equation}
\label{r8(3.2)}
\frac{1}{\lambda_{1}}\equiv\left[\ln^{2}\left(2a\right)
+\ln^{2}\left(\frac{\omega_{0}}{b_{2}}\right)
-\frac{2}{Vb_{1}}-2\right]^{\frac{1}{2}},
\end{equation}
and $a\equiv 2e^{\gamma}/\pi\simeq 1.13$ ($\gamma$ is the Euler constant). In this case the isotope coefficient is small and can be calculated from: $\alpha=\frac{\lambda_{1}}{2}\ln\left(\frac{\omega_{0}}{b_{2}}\right)$.

Returning to the central line of the thought, it can be easily seen, that for both regions of $U^{\left(0\right)}_{MF}$ the values of $\Delta_{tot}$ strongly increase when $U^{\left(0\right)}_{MF}$ increases. 
For $U^{\left(0\right)}_{MF}>U^{\left(0\right)}_{C}$ the evolution of the gap parameter with the temperature is sharply different from the classical BCS prediction (see Fig. \fig{f4}). In particular for $0<T<T_{C}$, the superconducting gap is very weak temperature dependent; this anomalous behavior is frequently observed in the cuprates \cite{Renner}. The another important results are presented for $T_{C} < T < T^{**}$, where $T^{**}$ denotes the highest value of the temperature for which the non-zero solution of the gap equation exists. In this case we have two branches. In order to find out, for which of these solutions the thermodynamic potential is lower, the numerical calculations have been made. The detailed analysis of this complicated issue is presented in the Appendix C. As an example, in Fig. \fig{f5} we show the dependence of the difference of the thermodynamic potential between the non-zero gap state and the normal state ($\Delta\Omega_{VU}$) on the temperature. The obtained result proves, that the physical solution represents the higher branch; whereas the lower branch corresponds to the unstable state. 

%
\begin{figure}[t]%
\includegraphics*[scale=0.15]{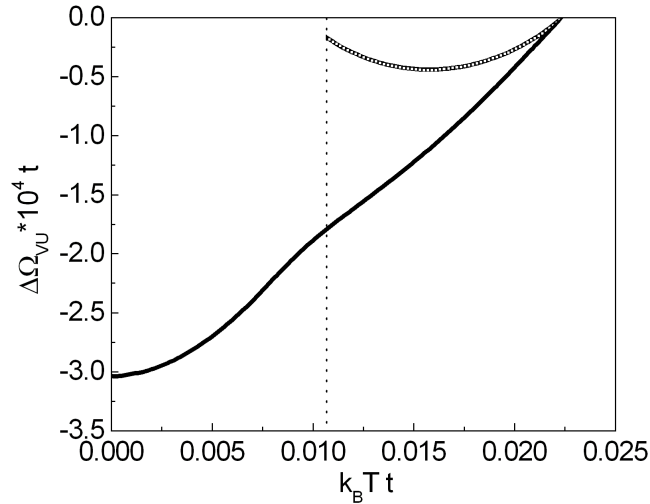}
\caption{The dependence of $\Delta\Omega_{VU}$ on the temperature for $U^{\left(0\right)}_{MF}=1.2370t$. We assume $V=1t$ and $\omega_{0}=0.3t$. The vertical line indicates the position of the critical temperature. The solid line for $0<T<T_{C}$ represents the result for the superconducting solution; the solid line for $T_{C} < T < T^{**}$ corresponds to the higher branch. The perforated line represents the thermodynamic potential for the lower branch. We notice that $\Delta\Omega_{VU}$ for the lower branch has the jump at $T_{C}$.  This behavior is connected with the inversion of the solutions of Eq. \eq{r5(3.2)} in the considered region viz., for $V_{1}$ and $U_{1}$ higher than $V_{2}$ and $U_{2}$ we have 
$\Delta_{tot}\left(V_{1},U_{1}\right)<\Delta_{tot}\left(V_{2},U_{2}\right)$ .}
\label{f5} 
\end{figure}
%

Next, the temperature $T^{**}$ should be interpreted on the experimental background.

First, we notice that the complicated mathematical structure of the order parameter (in general, it is the complex function) imposes two conditions on the existence of the superconducting state: (i) the amplitude of the order parameter has to differ from zero and  (ii) the superconducting state has to exhibit the long-range phase coherence. 

The essential pointer, how should be interpreted the temperature $T^{**}$, is connected with the experiments based on the Nernst effect \cite{Lee1}, \cite{Wang}. Namely, the Nernst signal above $T_{C}$ strongly suggests that the superconductivity vanishes at $T_{C}$ because the long-range phase coherence is destroyed by the thermally created vortices. Additionally, the experimental data have shown that the amplitude of the order parameter extends into the "normal" state to the temperature $T_{NE}$ (the Nernst temperature). We notice that $T_{NE}$ is considerably much lower than the pseudogap temperature ($T^{*}$). On the basis of presented experimental facts and the obtained theoretical results we assume that $T^{**}=T_{NE}$.

Before the comparison between the experimental data and the theoretical results obtained in the framework of toy model, we supplement the analytical approach. First, we notice that for $T=0$ the integral equation \eq{r5(3.2)} reduces to the algebraic equation:
\begin{equation}
\label{r9(3.2)}
\Delta_{tot}^{\left(0\right)}=2\omega_{0}e^{-\frac{1}{\lambda_{2}}},
\end{equation}
where the symbol $\Delta_{tot}^{\left(0\right)}$ denotes the gap parameter at zero temperature and
\begin{equation}
\label{r10(3.2)}
\frac{1}{\lambda_{2}}\equiv \ln\left(\frac{\omega_{0}}{b_{2}}\right)
+\left[\ln^{2}\left(\frac{\omega_{0}}{b_{2}}\right)-\frac{2}{V_{tot}b_{1}}
-\frac{\pi^{2}}{6}\right]^{\frac{1}{2}}.
\end{equation}

\begin{figure}[t]%
\includegraphics*[scale=0.15]{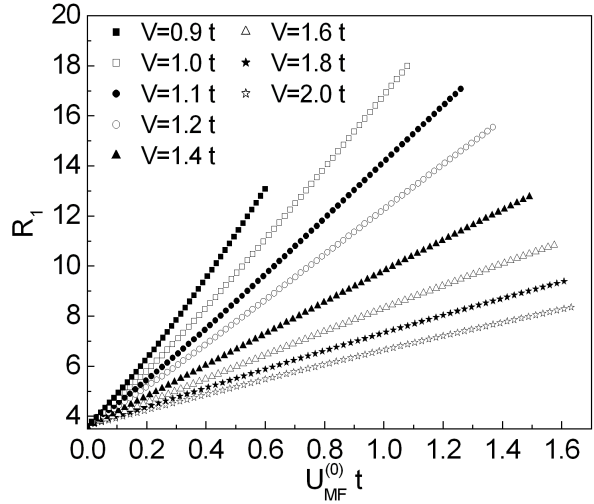}
\caption{The dependence of the ratio $R_{1}$ on $U^{\left(0\right)}_{MF}$ for different values of $V$. We assume $\omega_{0}=0.3t$. We notice that the ratio $R_{1}$ is plotted for different maximal values of $U^{\left(0\right)}_{MF}$, 
since this quantity is also strong dependent on $V$.}
\label{f6} 
\end{figure}

By using the equation \eq{r9(3.2)} and expression \eq{r7(3.2)} one can easily calculate the ratio: $R_{1}\equiv\frac{2\Delta_{tot}^{\left(0\right)}}{k_{B}T_{C}}$. In the classical BCS theory $R_{1}$ is the universal constant of the model and $\left[R_{1}\right]_{{\rm BCS}}=3.53$. In the case of the BCS van Hove scenario $R_{1}$ depends on the model parameters, however slightly ($\left[R_{1}\right]_{\rm max}\sim 4$).
The results obtained in the framework of the toy model are shown in Fig. \fig{f6}.
We see that $R_{1}$ is always bigger than in the BCS van Hove scenario and for sufficiently big values of  $U^{\left(0\right)}_{MF}$ the ratio $R_{1}$ achieves the physically acceptable values. 

%
%

Next, we consider the low-value behavior of the energy gap ($\Delta_{tot}/k_{B}T_{C}<<1$). In this case, the equation \eq{r5(3.2)} should be rewritten in the form:
\begin{equation}
\label{r11(3.2)}
1=\frac{2V_{tot}}{\beta}\sum^{+\infty}_{m=-\infty}\int^{\omega_{0}}_{0}
d\varepsilon\rho\left(\varepsilon\right)\frac{1}
{\omega^{2}_{m}+\varepsilon^{2}+\Delta^{2}_{tot}}.
\end{equation}
The kernel of the Eq. \eq{r11(3.2)} may be expanded in powers of $\Delta_{tot}$:
\begin{equation}
\label{r12(3.2)}
\frac{1}{\omega^{2}_{m}+\varepsilon^{2}+\Delta^{2}_{tot}}
\simeq\frac{1}{\omega^{2}_{m}+\varepsilon^{2}}-\frac{\Delta^{2}_{tot}}
{\left(\omega_{m}^{2}+\varepsilon^{2}\right)^{2}}.
\end{equation}
Next, we assume $\omega_{0}\rightarrow +\infty$, since $\omega_{0}>>k_{B}T_{C}$.
By using the lengthy but straightforward calculation we can transform the right-hand side of Eq. \eq{r11(3.2)} into the algebraic form:
\begin{equation}
\label{r13(3.2)}
1=b_{1}V_{tot}\left[p_{1}\left(T\right)-p_{2}\left(T\right)
\Delta^{2}_{tot}\right],
\end{equation}
where
\begin{equation}
\label{r14(3.2)}
p_{1}\left(T\right)\equiv\ln\left(\frac{2k_{B}T}{b_{2}}\right)
\ln\left(\frac{a\omega_{0}}{k_{B}T}\right)+\frac{1}{2}
\ln^{2}\left(\frac{\omega_{0}}{2k_{B}T}\right)-1,
\end{equation}
and
\begin{equation}
\label{r15(3.2)}
p_{2}\left(T\right)\equiv\left(\frac{1}{\pi k_{B}T}\right)^{2}
\left[\frac{7\zeta\left(3\right)}{8}\ln\left(\frac{\pi k_{B}T}{b_{2}e}\right)+\vartheta\left(3\right)\right].
\end{equation}
The symbol $\zeta$ denotes the Riemann zeta function, $\vartheta$ is defined by: $\vartheta\left(z\right)\equiv \sum^{+\infty}_{j=1}
\left(2j+1\right)^{-z}\ln\left(2j+1\right)$. 
If we omit the terms higher as $|\Delta|^{4}$ in Eq. \eq{r13(3.2)}, the expression for the anomalous thermal average takes the form:
\begin{equation}
\label{r16(3.2)} 
|\Delta^{\left(\pm\right)}|\simeq \sqrt{\left[r_{1}\left(T\right)
\pm\sqrt{r^{2}_{1}\left(T\right)+r_{2}\left(T\right)r_{3}\left(T\right)}\right]
r^{-1}_{2}\left(T\right)},
\end{equation}
where:
\begin{equation}
\label{r17(3.2)} 
r_{1}\left(T\right)\equiv \frac{1}{6}Up_{1}\left(T\right)-V^{3}p_{2}\left(T\right),
\end{equation}
\begin{equation}
\label{r18(3.2)} 
r_{2}\left(T\right)\equiv \frac{2}{3}UV^{2}p_{2}\left(T\right),
\end{equation}
and
\begin{equation}
\label{r19(3.2)} 
r_{3}\left(T\right)\equiv 2\left[p_{1}\left(T\right)V-\frac{1}{b_{1}}\right].
\end{equation}
We would like to point out that the expression \eq{r16(3.2)} can be used only in the framework of the generalized mean-field model where $U\neq 0$; in the case of the BCS van Hove scenario the formula \eq{r16(3.2)} generates the indeterminate form.

%
\begin{figure}[t]%
\includegraphics*[scale=0.15]{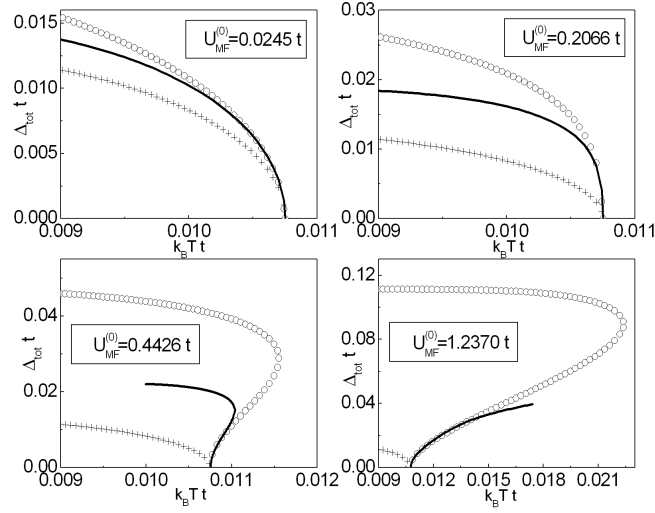}
\caption{The dependence of $\Delta_{tot}$ on the temperature close to the transition temperature for the small and large values of $U^{\left(0\right)}_{MF}$. We assume $V=1t$ and $\omega_{0}=0.3t$. The empty circles correspond to the exact numerical calculations (Eq. \eq{r5(3.2)}). The solid line represents the calculation of $\Delta_{tot}$ by using the formula \eq{r16(3.2)}. The crosses are obtained in the BCS van Hove scenario.
}
\label{f7} 
\end{figure}
%

In Fig. \fig{f7} we illustrate the temperature dependence of the gap parameter close to $T_{C}$ for different values of $U^{\left(0\right)}_{MF}$. In particular, we have shown the results obtained by using Eqs. \eq{r5(3.2)} and \eq{r16(3.2)}; for comparison we calculate also the gap parameter in the BCS van Hove scenario. One can see that the approximate formula \eq{r16(3.2)} very well reconstructs the exact numerical results both for the small and large values of $U^{\left(0\right)}_{MF}$; we strongly notice that for $U^{\left(0\right)}_{MF}>U^{\left(0\right)}_{C}$ exists only the unstable state. 

\subsection{The toy model and the experimental results}

The knowledge of the experimental values of  $T_{C}$, $T^{**}$ and other parameters ($t$ and $\omega_{0}$) enables the calculation of $V$ and $U^{\left(0\right)}_{MF}$ for the real materials. In particular, the value of $V$ is obtained by using the expression \eq{r7(3.2)}. Next, the potential $U^{\left(0\right)}_{MF}$ is determined from Eq. \eq{r5(3.2)}.
We notice that, in principle, the quantities  $V$ and $U^{\left(0\right)}_{MF}$ can be determined for the underdoped, optimally doped or overdoped samples; one can even take under consideration the influence of the disorder on $V$ and $U^{\left(0\right)}_{MF}$. This is possible, since the strength of the presented approach lies in fact that the physical state of the high-$T_{C}$ materials in the superconductivity domain is well reproduced by the values of $T_{C}$ and $T^{**}$.

%
\begin{table*}[h!]
\caption{\label{t1} The quantities $\delta^{**}$, $V$ and $U^{\left(0\right)}_{MF}$ calculated by using the mean values of $T_{C}$ and $T^{**}$. 
For YBCO the hole density $p$ is estimated from the relation presented in \cite{Liang}; for Bi2212 from the empirical formula $T_{C}\left(p\right)/T_{C,{\rm max}}=1-82.6\left(p-0.16\right)^{2}$ \cite{Presland}. For the superconductors Ni-NdBCO and Bi2223 the values of $t$ and $\omega_{0}$ are unknown. In this case, we take $t$ and $\omega_{0}$ for YBCO and Bi2212 respectively.}
\begin{ruledtabular}
\begin{tabular}{cccccccccccc}
Material & Type &$t$ (meV)&Ref.&$\omega_{0}$ (meV)&Ref.&$T_{C}$ (K)&$T^{**}$ (K)&Ref.&$\delta^{**}$&V (t)&$U^{\left(0\right)}_{MF}$ (t)\\
& & & & & & & & & &\\
\hline
YBCO &$p=0.062$   &250&\cite{Nunner}&75&\cite{Bohnen}&18.6&150.2&\cite{Ong} &{\bf 7.08}&{\bf 0.838}&{\bf 2.847}\\
     &$p=0.079$   &   &             &  &             &45  &128.3&           &{\bf 1.85}&{\bf 1.148}&{\bf 2.045}\\
     &$p=0.107$   &   &             &  &             &60.5&91.8 &           &{\bf 0.52}&{\bf 1.297}&{\bf 1.171}\\
     &$p=0.116$   &   &             &  &             &64.1&84.9 &           &{\bf 0.32}&{\bf 1.330}&{\bf 0.969}\\
     &$p=0.120$   &   &             &  &             &66.5&87.4 &           &{\bf 0.31}&{\bf 1.352}&{\bf 0.974}\\
     &$p=0.138$   &   &             &  &             &80.6&104.4&           &{\bf 0.30}&{\bf 1.475}&{\bf 1.062}\\
     &$p=0.150$   &   &             &  &             &90  &105  &           &{\bf 0.17}&{\bf 1.554}&{\bf 0.898}\\
     &$p=0.176$   &   &             &  &             &92  &107  &\cite{Wang}&{\bf 0.16}&{\bf 1.571}&{\bf 0.903}\\
\hline
$\rm YBCO_{e}$&$p=0.098$&250&\cite{Nunner}&75&\cite{Bohnen}&57  &85  &\cite{Rullier}&{\bf 0.49}&{\bf 1.265}&{\bf 1.104}\\
              &$p=0.098$&   &             &  &             &45.1&83.1&          &{\bf 0.84}&{\bf 1.150}&{\bf 1.288}\\
              &$p=0.098$&   &             &  &             &24.2&75  &          &{\bf 2.10}&{\bf 0.915}&{\bf 1.564}\\
              &$p=0.157$&   &             &  &             &92.6&103 &          &{\bf 0.11}&{\bf 1.576}&{\bf 0.796}\\
              &$p=0.157$&   &             &  &             &79.5&97.1&          &{\bf 0.22}&{\bf 1.465}&{\bf 0.930}\\
              &$p=0.157$&   &             &  &             &48.6&82.5&          &{\bf 0.70}&{\bf 1.184}&{\bf 1.213}\\
\hline
Zn-YBCO &${\rm x=0.000}$    &250&\cite{Nunner}&75&\cite{Bohnen}&90&$104\pm 5$&\cite{Xu1} &{\bf 0.16}&{\bf 1.554}&{\bf 0.876}\\
        &${\rm x=0.005}$&  &&  &          &84&$96 \pm 5$&                 &{\bf 0.14}&{\bf 1.504}&{\bf 0.813}\\
        &${\rm x=0.010}$ &  &&  &          &79&$87 \pm 5$&                &{\bf 0.10}&{\bf 1.461}&{\bf 0.700}\\
        &${\rm x=0.020}$ &  &&  &          &67&$75 \pm 5$&                &{\bf 0.12}&{\bf 1.356}&{\bf 0.674}\\
\hline
Pr-YBCO &${\rm x=0.0}  $&250&\cite{Nunner}&75&\cite{Bohnen}&89.7&$104.8\pm 5$&\cite{Li2} &{\bf 0.17}&{\bf 1.552}&{\bf 0.899}\\
        &${\rm x=0.1}$&   &             &  &             &83.8&$99.9\pm 5$ &    &{\bf 0.19}&{\bf 1.502}&{\bf 0.908}\\
        &${\rm x=0.2}$&   &             &  &             &68.2&$95  \pm 5$ &    &{\bf 0.39}&{\bf 1.367}&{\bf 1.096}\\
        &${\rm x=0.3}$&   &             &  &             &50.2&$84.8\pm 5$ &    &{\bf 0.69}&{\bf 1.200}&{\bf 1.226}\\
        &${\rm x=0.4}$&   &             &  &             &40.7&$79.9\pm 5$ &    &{\bf 0.96}&{\bf 1.104}&{\bf 1.313}\\
\hline
Ni-NdBCO & x=0.00,y=0.0   &   250   & -  &     75 &  - & 95         &$115\pm 20$\footnote{The high value of the experimental error.} &\cite{Johannsen}&{\bf 0.21}&{\bf 1.596}&{\bf 1.021}\\
         & x=0.03,y=0.0   &         &    &        &    &$59\pm 4$   &$80\pm 20^{\it a}$  &                &{\bf 0.36}&{\bf 1.283}&{\bf 0.966}\\
         & x=0.06,y=0.0   &         &    &        &    &$45\pm 5$   &$65 \pm 20^{\it a}$ &                &{\bf 0.44}&{\bf 1.148}&{\bf 0.940}\\
         & x=0.00,y=0.2   &         &    &        &    &$53\pm 3.5$ &$75\pm 5$   &                &{\bf 0.42}&{\bf 1.227}&{\bf 0.981}\\
         & x=0.03,y=0.2   &         &    &        &    &$20\pm 7$   &$80\pm 10$  &                &{\bf 3.00}&{\bf 0.858}&{\bf 1.737}\\
\hline
Bi2212 &$p=0.087$&$350$&\cite{Tohayama},\cite{Tohayama1},&$80$&\cite{Damascelli},\cite{Cuk},&$50$&$108.9\pm 5$&\cite{Wang},\cite{Wang1}.
                                                                                &{\bf 1.18}&{\bf 1.117}&{\bf 1.508}\\
        &$p=0.118$&     &\cite{Kim1}.&    &\cite{Gweon},\cite{Kulic1},&$77.9$&$130.3\pm 10$&  &{\bf 0.67}&{\bf 1.347}&{\bf 1.442}\\
        &$p=0.160$&     &            &    & \cite{Gonnelli}.       &$90.6$&$125.4\pm 5$ &  &{\bf 0.38}&{\bf 1.444}&{\bf 1.184}\\
        &$p=0.202$&     &  &    &               &$76.9$&$105.5\pm 5$ &                     &{\bf 0.37}&{\bf 1.339}&{\bf 1.050}\\
        &$p=0.219$&     &  &    &               &$64.9$&$85.5 \pm 10$&                     &{\bf 0.32}&{\bf 1.243}&{\bf 0.888}\\
\hline
Bi2223 & OP & 350  & - &  80 & - & $109$ & $135$ & \cite{Wang} & {\bf 0.24} & {\bf 1.580} & {\bf 1.077}\\
\hline
PCCO &${\rm x=0.13}$&380&\cite{Andersen},\cite{Lin},&33&\cite{Khlopkin},\cite{Balci}.&$12  \pm 1$&$18  \pm 1$&\cite{Li}
                                                                             &{\bf 0.50}&{\bf 0.795}&{\bf 0.715}\\
     &${\rm x=0.14}$&  &\cite{Zimmers},\cite{Hackl}.&  &  &$19.5\pm 1$&$23  \pm 1$&             &{\bf 0.18}&{\bf 0.947}&{\bf 0.529}\\
     &${\rm x=0.15}$&  &                            &  &  &$20  \pm 1$&$22  \pm 1$&             &{\bf -}   &{\bf 0.956}& {\bf 0.174}\footnote{The value found based on $\left[R_{1}\right]_{{\rm exp}}$.}\\
     &${\rm x=0.17}$&  &                            &  &  &$13  \pm 1$&$13.5\pm 1$&             &{\bf -}   &{\bf 0.817}&${\bf 0.002}^{\it b}$\\
     &${\rm x=0.19}$&  &                            &  &  &$6.5 \pm 1$&$7   \pm 1$&             &{\bf -}   &{\bf 0.654}&${\bf \sim 0}^{\it b}$
\end{tabular}
\end{ruledtabular}
\end{table*}
%

In the subsection (for selected superconductors), on the basis of  $V$ and $U^{\left(0\right)}_{MF}$, we calculate the dependence of the ratio $R_{1}$ on the hole density or on the doping; we consider also the influence of the disorder on $R_{1}$. We compare the theoretical predictions with the experimental results (if the experimental data is known). In one case we plot openly the dependence of the energy gap on the temperature, since the right experimental data has existed in the literature.    

In particular, we take under consideration following families of the high-$T_{C}$ materials: 
${\rm YBa_{2}Cu_{3}O_{7-y}}$ (YBCO) - for selected values of the hole density ($p$ - (holes/Cu)), 
${\rm YBa_{2}Cu_{3}O_{7-y}}$ ($\rm YBCO_{e}$) - the disorder induced by electron irradiation, which results in the creation of point defects such as Cu and O vacancies in the $\rm CuO_{2}$ planes,
${\rm YBa_{2}\left(Cu_{1-x}Zn_{x}\right)_{3}O_{7-y}}$ (Zn-YBCO) - the in-plane disorder caused by zinc,
${\rm Y_{1-x}Pr_{x}Ba_{2}Cu_{3}O_{7-y}}$ (Pr-YBCO) - the out-of-plane disorder caused by praseodymium,
${\rm NdBa_{2}(Cu_{1-x}Ni_{x})_{3}O_{7-y}}$ (Ni-NdBCO) - for selected values of x and y.  
Next, we consider ${\rm Bi_{2}Sr_{2}CaCu_{2}O_{8+y}}$ (Bi2212) - for selected values of the hole density, and 
${\rm Bi_{2}Sr_{2}Ca_{2}Cu_{3}O_{10+y}}$ (Bi2223) - for the optimally doped case (OP). Finally, 
${\rm Pr_{2-x}Ce_{x}CuO_{4-y}}$ (PCCO) - the example of electron-doped cuprate.

The experimental data and the obtained theoretical results are collected in the Tab. \tab{t1}.
Mathematically, the range of the Nernst region can be characterized by the quantity:
$\delta^{**}\equiv\left(T^{**}-T_{C}\right)/T_{C}$. On the basis of experiments, we conclude that the Nernst region strongly expands if the hole density decreases; this is clear to see for YBCO and Bi2212, where $\left[\delta^{**}\right]_{\rm max}=7.08$ and  $\left[\delta^{**}\right]_{\rm max}=1.18$ respectively; also for underdoped Ni-NdBCO, where $\left[\delta^{**}\right]_{\rm max}=3$. In the case of the disorder which is induced in YBCO, the value of $\delta^{**}$ considerably increases if the electron irradiation is used or yttrium is substituted by praseodymium; the disorder caused by zinc weakly influences on the value of $\delta^{**}$. In the presented analysis we consider also the electron-doped cuprate system PCCO for which $\delta^{**}$ increases if doping decreases (from optimally to underdoped region). In overdoped region of PCCO the value of $\delta^{**}$ can not be determined, since  $T_{C}$ and $T^{**}$ are experimentally indistinguishable.
 
Now, we consider the obtained values of $V$ and $U^{\left(0\right)}_{MF}$. We notice that for real materials both $V$ and $U^{\left(0\right)}_{MF}$ are significant and the following principle is in effect: if $\delta^{**}$ is smaller than $\sim 0.6$  we have $V>U^{\left(0\right)}_{MF}$, in the opposite case $V<U^{\left(0\right)}_{MF}$; the especially high values of $U^{\left(0\right)}_{MF}$ can be observed in the strongly underdoped regime.

By using the input parameters presented in Tab. \tab{t1} we compare the calculated theoretical values of the ratio $R_{1}$ with the experimental data. 

\begin{figure}[t]%
\includegraphics*[scale=0.15]{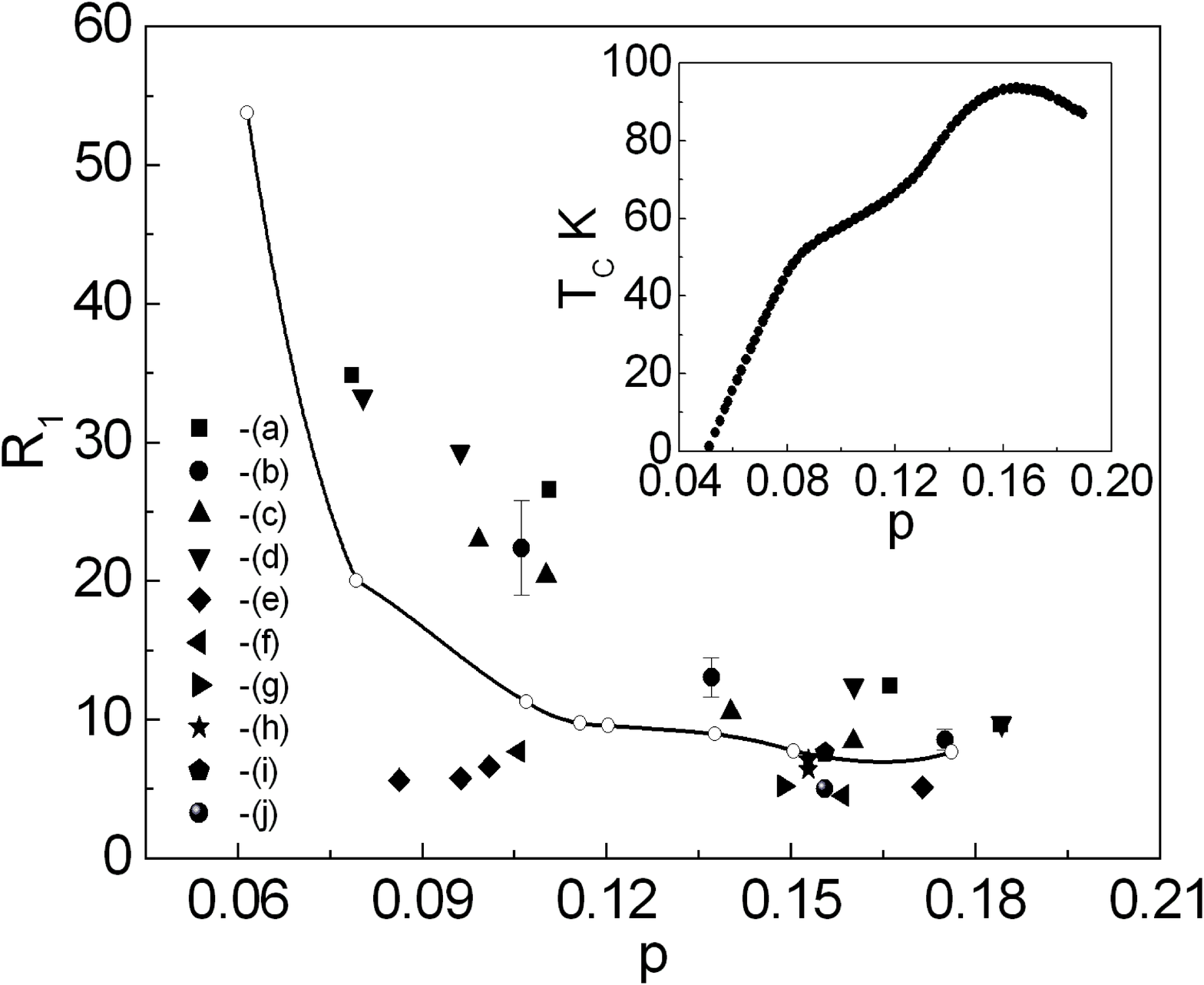}
\caption{The dependence of the ratio $R_{1}$ on $p$ for YBCO. The solid line with the open circles represents the theoretical calculation. The black filled symbols correspond to experimental results obtained by: 
(a) - M. Sutherland {\it et al.} \cite{Sutherland}, 
(b) - K. Nakayama {\it et al.} \cite{Nakayama}, 
(c) - A. Kaminski {\it et al.} \cite{Kaminski}, 
(d) - M. Plate {\it et al.} \cite{Plate},
(e) - D.K. Morr {\it et al.} \cite{Morr}, H.F. Fong {\it et al.} \cite{Fong},
(f) - N.C. Yeh {\it et al.} \cite{Yeh},
(g) - V. Born {\it et al.} \cite{Born},
(h) - H. Murakami {\it et al.} \cite{Murakami},
(i) - H. Edwards {\it et al.} \cite{Edwards1},
(j) - H. Edwards {\it et al.} \cite{Edwards2}. 
The inset shows the dependence of the critical temperature on the hole density estimated from the relation presented in \cite{Liang}.}
\label{f8} 
\end{figure}

In Fig. \fig{f8} we show the ratio $R_{1}$  as a function of the hole density for YBCO. The numerical results are denoted by the solid line with the open circles; the experimental data by the black filled symbols (see also the Tab. \tab{t2} in Appendix D). It can be seen that, from slightly underdoped to overdoped crystals the presented model correctly predicts the values of $R_{1}$ (taking under consideration the several approximations which have already been mentioned previously). For $p\in\left(0.07,0.135\right)$ the theoretical values of  the ratio $R_{1}$ are probably lower than the experimental data. However, the increasing of $R_{1}$ with decreasing of the hole concentration is qualitatively correctly predicted. In the case of the strongly underdoped crystals ($p<0.07$) the toy model suggests very high values of the ratio $R_{1}$ which have to be experimentally checked. In the inset in Fig. \fig{f8} there is presented the open dependence of the critical temperature on the hole density with help of which the values of $p$ for YBCO have been calculated \cite{Liang}.
 
\begin{figure}[t]%
\includegraphics*[scale=0.15]{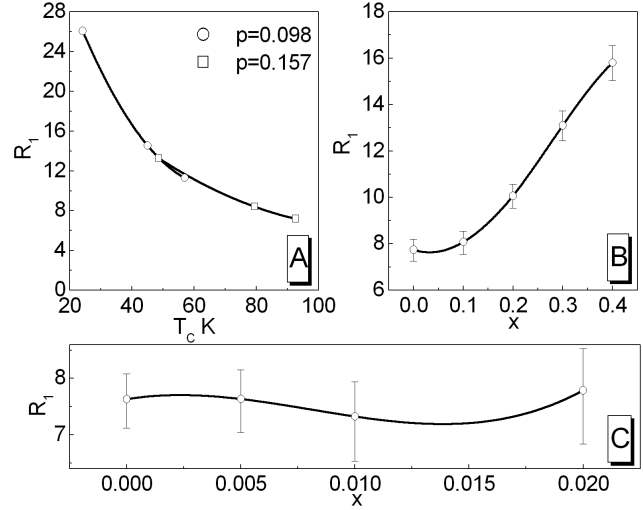}
\caption{
({\it A}) The dependence of the ratio $R_{1}$ on $T_{C}$ for $\rm YBCO_{e}$; we consider the case $p=0.098$ and $p=0.157$. 
({\it B}) The dependence of the ratio $R_{1}$ on x for Pr-YBCO.
({\it C}) The dependence of the ratio $R_{1}$ on x for Zn-YBCO.}
\label{f9} 
\end{figure}

In Figs. \fig{f9} (A)-(C) for YBCO we presented the influence of the disorder on the value of the ratio $R_{1}$. In Fig. \fig{f9} (A), the dependence of $R_{1}$ on $T_{C}$ for the disorder induced by electron irradiation is shown. The two values of the hole concentration are considered: $p=0.098$ and $p=0.157$. In both cases the growing disorder, which is being manifested by the lowering values of $T_{C}$, causes very distinct increase of the ratio $R_{1}$. It is possible to observe the similar effect for the case, when the out-of-plane disorder is caused by praseodymium (Fig. \fig{f9} (B)). However, the in-plane disorder which is induced by zinc in principle doesn't affect the value of the parameter $R_{1}$; see Fig. \fig{f9} (C). In Figs. \fig{f9} (A)-(C) we have shown only the theoretical predictions, since the experimental data not yet exist.
%
%
%
%

\begin{figure}[t]%
\includegraphics*[scale=0.15]{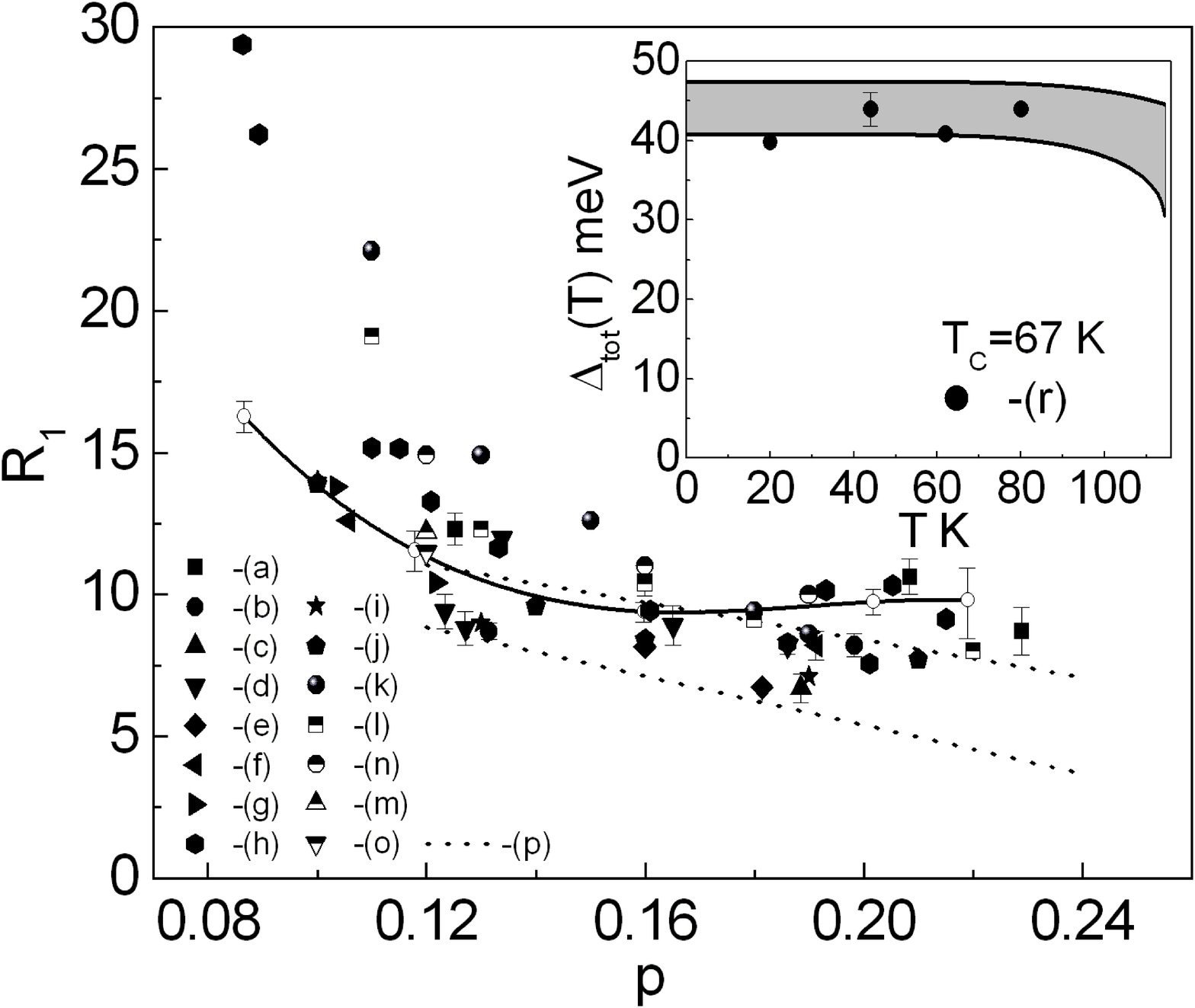}
\caption{The dependence of the ratio $R_{1}$ on $p$ for Bi2212. The solid line with the open circles represents the theoretical calculation. The filled and half-filled symbols correspond to experimental results obtained by:
(a) - Ch. Renner, {\it et al.} \cite{Renner}, 
(b) - A. Hoffmann, {\it et al.} \cite{Hoffmann},
(c) - Y.G. Ponomarev, {\it et al.} \cite{Ponomarev}, 
(d) - T. Oki, {\it et al.} \cite{Oki},
(e) - V.M. Krasnov, {\it et al.} \cite{Krasnov}, 
(f) - A.K. Gupta, {\it et al.} \cite{Gupta},
(g) - A. Kanigel, {\it et al.} \cite{Kanigel}, 
(h) - J.C. Campuzano, {\it et al.} \cite{Campuzano}, K. Tanaka, {\it et al.} \cite{Tanaka},
(i) - T. Nakaono, {\it et al.} \cite{Nakaono},
(j) - M. Oda,     {\it et al.} \cite{Oda},
(k) - K. McElroy, {\it et al.} \cite{McElroy},
(l) - A. Matsuda, {\it et al.} \cite{Matsuda},
(m) - J.E. Hoffman, {\it et al.} \cite{Hoffman},
(n) - C. Howald, {\it et al.} \cite{Howald},
(o) - H. Murakami, {\it et al.} \cite{Murakami1}.
The lines (p) are obtained by using the empirical relation \eq{r1(3.3)}.
The inset shows the dependence of the energy gap on the temperature; the filled region between the solid lines represents possible theoretical values of $\Delta_{tot}$; (r) - the experimental results \cite{Kanigel}.
}
\label{f10} 
\end{figure}

The dependence of the ratio $R_{1}$ on the hole density for Bi2212 is presented in Fig.\fig{f10}. Similarly as for YBCO, the theoretical results are denoted by the solid line with the open circles and the experimental data by the filled and half-filled symbols (the Tab. \tab{t4} in Appendix D). 
Additionally, the region between the dotted lines represents the possible experimental 
values obtained by using the empirical formula \cite{Hewitt}:
\begin{equation}
\label{r1(3.3)} 
R_{1}\left(p\right)=\left(15\pm 1\right)-\left(38\pm 5\right)p. 
\end{equation}
We can notice that relation \eq{r1(3.3)} represents the linear least-squares fit to the experimental data and it is valid for $0.12<p<0.24$.
On the basis of the presented results we conclude, that the toy model, in a wide range of the $p$ values, very correctly reproduces the experimental data. It is possible to have reservation only for very low values of $p$, where some experimental data suggested extremely high values of the ratio $R_{1}$. The inset in Fig. \fig{f10} presents the dependence of the energy gap on the temperature for Bi2212 with $T_{C}=67$ K. In particular, the region between the solid lines corresponds to the possible theoretical values of $\Delta_{tot}$. Let us notice that theoretical results were set with the experimentally accuracy of $T^{**}$ (see also the Tab. \tab{t1}). In the inset we also show the experimental data obtained by A. Kanigel, {\it et al.} \cite{Kanigel}. It is easy to see that the agreement between theoretical and experimental results is excellent.

%
%

However, for Bi2223 and Ni-NdBCO superconductors the values of $t$ and $\omega_{0}$ are unknown and we take $t$ and $\omega_{0}$ for Bi2212 and YBCO   respectively, it is possible to receive good estimating of the real values of $R_{1}$. In the case of Bi2223 we have $R_{1}=8.01$. On the basis of the results presented in Appendix D (Tab.\tab{t5}) we see that our theoretical value is very close to the experimental data. For Ni-NdBCO the situation is more complicated. If we consider Ni-NdBCO superconductor with y=0 the experimental error for $T^{**}$ is too big, so it is impossible exactly determine $R_{1}$. However, if we take the mean values of $T_{C}$ and $T^{**}$ the theoretical results indicate that the ratio $R_{1}$ has the values $7.97$, $10.20$, $11.78$ respectively for x=0, x=0.03 and x=0.06. We notice that the experimental value of $R_{1}$ for x=0 is equal to $7.3$ (see the 
Tab.\tab{t3} in Appendix D). For the case y=0.2 the experimental data are much more accurate and we have $R_{1}$ equal to $10.98^{+1.69}_{-1.65}$ and $32.90^{+21.43}_{-10.68}$ for x=0 and x=0.03 respectively. On the basis of above results we see that the last value of the ratio $R_{1}$ can be extraordinary big (this result should be checked experimentally).

The general phase diagram of the high-$T_{C}$ superconductors presents the global symmetry between the hole- and electron-doped materials \cite{Almasan}. First of all, in both cases, the antiferromagnetic phase has the similar Neel temperature (however for electron-doped cuprates, the antiferromagnetic phase is broader). Secondly, the superconducting phase for the hole- and electron-doped materials appears closely to antiferromagnetic phase, with the similar value of the optimal doping (${\rm x}\sim 0.16$). The distinct symmetry of the phase diagram strongly supports the view that the hole- and electron-doped superconductors should have an identical pairing mechanism. For that reason, the analysis of $R_{1}$ for the selected electron-doped superconductor in the framework of toy model is very important. 

\begin{figure}[t]%
\includegraphics*[scale=0.15]{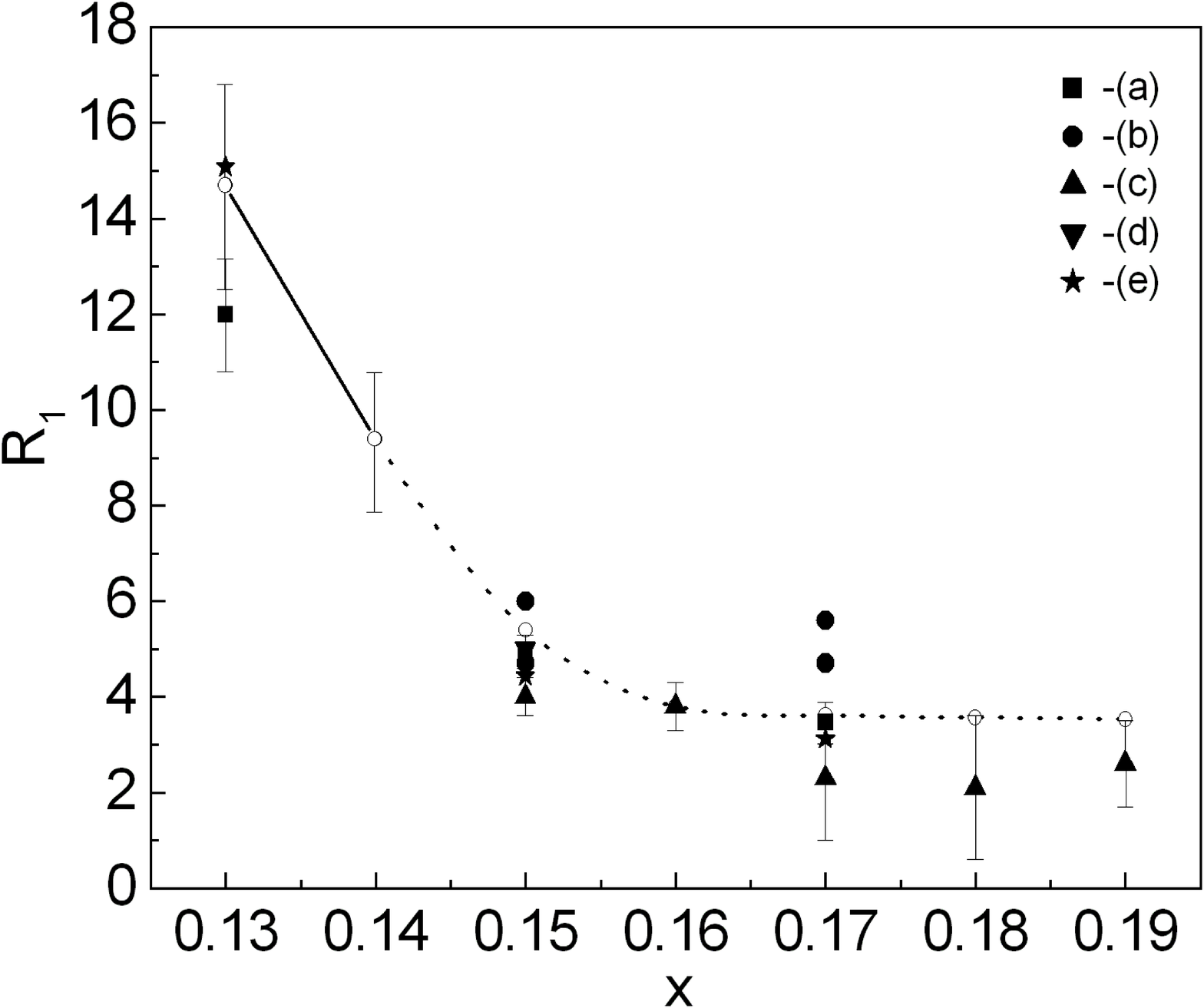}
\caption{The dependence of the ratio $R_{1}$ on x for PCCO. The solid line with the open circles represents the theoretical calculation based on $T_{C}$ and $T^{**}$; the dotted line with open circles represents the theoretical calculation based on $T_{C}$ and the appropriately selected $U^{\left(0\right)}_{MF}$. The black filled symbols correspond to experimental results obtained by: 
(a) - A. Biswas,   {\it et al.} \cite{Biswas}, 
(b) - A. Zimmers,  {\it et al.} \cite{Zimmers1},
(c) - Y. Dagan,    {\it et al.} \cite{Dagan}, 
(d) - C.C. Homes,  {\it et al.} \cite{Homes},
(e) - P. Fournier, {\it et al.} \cite{Fournier}.
}
\label{f11} 
\end{figure}

In Fig. \fig{f11} we show the dependence of the ratio $R_{1}$ on doping for PCCO. The solid line with open circles represents the theoretical calculation obtained by using the input parameters $T_{C}$ and $T^{**}$; the dotted line with open circles represents the theoretical calculation obtained by using the input parameters $T_{C}$ and the appropriately selected $U^{\left(0\right)}_{MF}$. The black filled symbols correspond to experimental data. Based on Fig. \fig{f11}, it is easy to see that the theoretical results very well reconstruct the experimental values of $R_{1}$.

%
%

\section{SUMMARY}

In the study the simple microscopic model of the superconducting state that induce at high temperature was presented. The starting point of our considerations was the assumed statement: that a proper description of the superconducting condensate in the cuprates is possible only when the pairing mechanism would {\it inseparably} link together the strong electron correlations and the crystal lattice vibrations. The Hamiltonian proposed in the paper seems to be the most simple among the operators that satisfy the above postulate. 

In the study we have shown, that for the large values of the electron-electron-phonon coupling, the energy gap at the vicinity of the superconducting state existence weakly depends on the temperature and it vanishes above $T_{C}$ (at the temperature that was interpreted as Nernst temperature). The key test of our model was the determination of the dependence of the ratio $R_{1}$ on doping for the selected superconductors. The obtained agreement between the theoretical and experimental data seems to be extraordinary, when taking into account the simplicity of the considered Hamiltonian and used approximations. 

In our opinion, the achieved results clearly suggest that presented pairing mechanism should be seriously taken into consideration in the further researches which may lead to the most precise understanding of the properties of the high temperature superconducting state.

\begin{acknowledgments}
The author wish to thank Prof. K. Dzili{\'n}ski, the Head of the Institute of Physics at Cz{\c{e}}stochowa University of Technology, for providing excellent working conditions and the financial support. Additionally, I would like to thank my colleagues and students: D. Szcz{\c{e}}{\'s}niak, M.W. Jarosik, A.P. Durajski and T. Mila for their kindness and technical support given during the preparation of this work. Some computational resources have been provided by the RSC Computing Center.
\end{acknowledgments}

\appendix
\section{The exact expression for the self-energy matrix}

In the framework of the Eliashberg formalism, the Nambu spinors are defined in the following way:
\begin{equation}
\label{D1}
\Psi _{\kvec}\equiv\left( 
\begin{array}{l}
c_{\kvec\uparrow } \\ 
c_{-\kvec\downarrow}^{\dagger}
\end{array}
\right) ,\Psi _{\kvec}^{\dagger}\equiv\left( 
\begin{array}{ll}
c_{\kvec\uparrow }^{\dagger} & c_{-\kvec\downarrow}
\end{array}
\right), 
\end{equation}
where $\Psi _{\kvec}$ and $\Psi _{\kvec}^{\dagger}$ satisfy the anticommutation rules:
\begin{equation}
\label{D2}
\left[\Psi_{\kvec},\Psi_{\kvec^{'}}^{\dagger}\right]_{+}=
\delta_{\kvec \kvec^{'}}\tau_{0}.
\end{equation}

In the presented notation, the terms of the Hamiltonian \eq{r1(2.0)} can be rewritten as follows:
\begin{equation}
\label{D3}
H^{\left(0\right)}=\sum_{\kvec}\overline\varepsilon_{\kvec}\Psi _{\kvec}^{\dagger}\tau_{3}\Psi _{\kvec}+\sum_{\qvec}\omega_{\qvec}b^{\dagger}_{\qvec}b_{\qvec},
\end{equation}
\begin{equation}
\label{D4}
H^{\left(1\right)}=\sum_{\kvec}
g^{\left(1,2\right)}_{\kvec}\left(\bf{q}\right)\Psi _{\kvec+\qvec}^{\dagger}\tau_{3}\Psi _{\kvec}\phi_{\qvec},
\end{equation}
and
\begin{widetext}
\begin{equation}
\label{D5}
H^{\left(2\right)}=\sum_{\kvec\kvec^{'}\qvec\lvec}
g^{\left(2a\right)}_{\kvec,\kvec^{'}}\left({\bf q},{\bf l}\right) 
\Psi_{\kvec}^{\dagger}\tau _{B}\Psi _{-\kvec^{'}}
\Psi_{-\kvec^{'}-\lvec+\qvec}^{\dagger}\tau_{C}\Psi_{\kvec-\lvec}\phi_{\qvec}
+\sum_{\kvec\kvec^{'}\qvec\lvec}
g^{\left(2b\right)}_{\kvec,\kvec^{'}}\left({\bf q},{\bf l}\right) 
\Psi_{\kvec}^{\dagger}\tau _{C}\Psi _{-\kvec^{'}}
\Psi_{-\kvec^{'}-\lvec+\qvec}^{\dagger}\tau_{B}\Psi_{\kvec-\lvec}\phi_{\qvec}.
\end{equation}
\end{widetext}
The matrix elements take the form:
\begin{equation}
\label{D6} 
g^{\left(1,2\right)}_{\kvec}\left(\bf{q}\right)\equiv g^{\left(1\right)}_{\kvec}\left(\bf{q}\right)+ g^{\left(2\right)}_{\kvec}\left(\bf{q}\right),
\end{equation}
where: 
\begin{equation}
\label{D7}
g^{\left(2\right)}_{\kvec}\left(\bf{q}\right)
\equiv\sum_{\kvec^{'}}g^{\left(2\right)}_{\kvec^{'},\kvec}\left({\bf q},0\right)=\sum_{\kvec^{'}}g^{\left(2\right)}_{\kvec^{'},\kvec}\left({\bf q},-\bf{q}\right),
\end{equation}
and
\begin{equation}
\label{D8}
g^{\left(2x\right)}_{\kvec,\kvec^{'}}\left({\bf q},{\bf l}\right)\equiv
\left\{ 
\begin{array}{lll}
g^{\left(2\right)}_{\kvec-\lvec,\kvec^{'}+\lvec-\qvec}\left({\bf q},{-\bf l}\right) & \rm{for} & x=a \\ 
g^{\left(2\right)}_{-\kvec,-\kvec^{'}}\left({\bf q},{-\bf l}\right)& \rm{for} & x=b.\\ 
\end{array}\right.
\end{equation}
The bases of the matrices $\tau_{A}$-$\tau_{D}$ and $\tau_{0}-\tau_{3}$ (the Pauli matrices) are presented below: 
\begin{eqnarray}
\label{D9}
\tau_{A}&\equiv&\left(\begin{array}{ll}
1 & 0\\ 
0 & 0 
\end{array}\right)=\frac{1}{2}\left(\tau_{0}+\tau_{3}\right),\\
\tau_{B}&\equiv&\left(\begin{array}{ll}
0 & 1\\ 
0 & 0 
\end{array}\right)=\frac{1}{2}\left(\tau_{1}+i\tau_{2}\right),\\
\tau_{C}&\equiv&\left(\begin{array}{ll}
0 & 0\\ 
1 & 0 
\end{array}\right)=\frac{1}{2}\left(\tau_{1}-i\tau_{3}\right),\\
\tau_{D}&\equiv&\left(\begin{array}{ll}
0 & 0\\ 
0 & 1 
\end{array}\right)=\frac{1}{2}\left(\tau_{0}-\tau_{3}\right).
\end{eqnarray}

With the help of definition \eq{D1} we have introduced the electron Green function 
$G_{\kvec}(i\omega _n)\equiv \left\langle \left\langle \Psi _{\kvec}|\Psi
_{\kvec}^{\dagger}\right\rangle \right\rangle_{i\omega_{n}}$, where: $\omega_{n}\equiv \left(\pi / \beta\right)\left(2n-1\right)$ is the $n$-th Matsubara frequency and $\beta\equiv\left(k_{B}T\right)^{-1}$ ($k_{B}$ is the Boltzmann constant). 
The electron propagator $G_{\kvec}(i\omega _n)$ can be written in the form: 
\begin{equation}
\label{D10} 
G_{\kvec}(i\omega _n)=
\left( \begin{array}{cc}
<< c_{\kvec\uparrow}|c_{\kvec\uparrow}^{\dagger}>>_{i\omega_{n}}&
<< c_{\kvec\uparrow}|c_{-\kvec\downarrow}>>_{i\omega_{n}} \\
<< c_{-\kvec\downarrow}^{\dagger}|c_{\kvec\uparrow}^{\dagger}>>_{i\omega_{n}}&
<< c_{-\kvec\downarrow}^{\dagger}|c_{-\kvec\downarrow}>>_{i\omega_{n}}
\end{array}\right).
\end{equation}
We notice that the superconducting thermal average should be evaluated from the non-diagonal part of $G_{\kvec}(i\omega_n)$, whereas the diagonal part determines the normal-state properties.
It is essential, that the Hamiltonian \eq{r1(2.0)} in the matrix form enables to obtain the Dyson equation:
\begin{equation}
\label{D11}
G_{\kvec}(i\omega _n)=G_{0\kvec}(i\omega _n)+G_{0\kvec}(i\omega _n)M_{\kvec}(i\omega _n)G_{0\kvec}(i\omega _n),
\end{equation}
where $G_{0\kvec}(i\omega _n)$ denotes the unperturbed ($g^{\left(1\right)}_{\kvec}\left({\bf q}\right)=g^{\left(2\right)}_{\kvec,\kvec^{'}}\left({\bf q},{\bf l}\right)=0$) propagator:
\begin{equation}
\label{D12}
G_{0\kvec}(i\omega _n)\equiv\left( i\omega _n\tau _0-\overline\varepsilon\tau_{3}\right) ^{-1},
\end{equation}
and the self-energy matrix has the form:
\begin{widetext} 
\begin{eqnarray}
\label{D13} 
M_{\kvec}(i\omega _n)&\equiv&
\left<\left[\left[\Psi_{\kvec},H^{\left(2\right)}\right]_{-},
\Psi^{\dagger}_{\kvec}\right]_{+}\right>\\ \nonumber
&+&
<< 
\sum_{\qvec}g^{\left(1,2\right)}_{\kvec-\qvec}\left(\bf{q}\right)\tau_{3}\Psi _{\kvec-\qvec}\phi _{\qvec}
|
\sum_{\qvec^{'}}g^{\left(1,2\right)}_{\kvec}\left(\bf{q^{'}}\right) 
\Psi _{\kvec+\qvec^{'}}^{\dagger}\phi _{\qvec^{'}}\tau_{3}
>>_{i\omega_{n}}\\ \nonumber
&+&
<< 
\sum_{\qvec}g^{\left(1,2\right)}_{\kvec-\qvec}\left(\bf{q}\right)\tau_{3}\Psi _{\kvec-\qvec}\phi _{\qvec}
|
\left[H^{\left(2\right)}, \Psi^{\dagger}_{\kvec}\right]_{-}
>>_{i\omega_{n}}\\ \nonumber
&+&
<< 
\left[\Psi_{\kvec},H^{\left(2\right)}\right]_{-}
|
\sum_{\qvec^{'}}g^{\left(1,2\right)}_{\kvec}\left(\bf{q^{'}}\right) 
\Psi _{\kvec+\qvec^{'}}^{\dagger}\phi _{\qvec^{'}}\tau_{3}
>>_{i\omega_{n}}\\ \nonumber
&+&
<< 
\left[\Psi_{\kvec},H^{\left(2\right)}\right]_{-}
|
\left[H^{\left(2\right)}, \Psi^{\dagger}_{\kvec}\right]_{-}
>>_{i\omega_{n}}. 
\end{eqnarray}
\end{widetext}

The equation \eq{D13} represents the formally exact expression for the self-energy matrix. The first-order contribution to $M_{\kvec}(i\omega _n)$ comes from the thermal average $\left<\right>$ and the second-order contribution from the propagators in angular brackets. 
The complete analysis of the matrix self-energy is very difficult problem. However, one can demonstrate that, in the case of the lattice distortion's absence ($\left<b_{\qvec}\right>=0$), the first-order contribution to the self-energy is equal to zero. In particular: 

\begin{equation}
\label{D14}  
\left<\left[\left[\Psi_{\kvec},H^{\left(2\right)}\right]_{-},\Psi^{\dagger}_{\kvec}\right]_{+}\right>=
\left<\left[\left[\Psi_{\kvec},\widetilde{H}^{\left(2\right)}\right]_{-},\Psi^{\dagger}_{\kvec}\right]_{+}\right>
\left<b_{-\qvec}^{\dagger}+b_{\qvec}\right> =0,
\end{equation}
where $\widetilde{H}^{\left(2\right)}$ represents the fermionic parts in the Hamiltonian $H^{\left(2\right)}$. We notice that the non-diagonal part of the expression \eq{D14} is proportional to the superconducting thermal average, whereas the diagonal part is connected with the average number of electrons per lattice site. 

The standard electron-phonon self-energy is represented by the second term in Eq. \eq{D13}, which is proportional to $\left(g^{\left(1,2\right)}_{\kvec}\left(\bf{q}\right)\right)^{2}$. On the basis of Eqs. \eq{D6} and \eq{D7} it is easy to see that the value of the electron-phonon coupling can be directly increased by the electron-electron-phonon interaction. One should clearly emphasize the fact that the existence of the  $g^{\left(2\right)}_{\kvec}\left(\bf{q}\right)$ element is the direct consequence of the form of the Hamiltonian \eq{r1(2.0)} in the symmetric Nambu notation. From the physical point of view, the above result is highly nontrivial. On the other hand, the remaining terms also give the important  contributions to the electron-phonon pairing and, on the Eliashberg equations level, they can not be neglected.      

\section{The {\it fold} mean-field approximation of the 4EE Hamiltonian}

We have rewritten the interaction term in the Hamiltonian \eq{r6(3.1)} in the form:  
\begin{equation}
\label{A1}
H_{int}\equiv-\frac{U}{24N^{3}}
\sum^{\omega_{0}}_{\kvec\kvec^{'}\qvec\lvec\sigma}
c_{\kvec-\lvec\sigma }^{\dagger}c_{\kvec\sigma}
h^{\left(1\right)}_{\kvec^{'}\lvec\qvec\sigma}
c_{-\kvec+\lvec-\sigma }^{\dagger}c_{-\kvec-\sigma},
\end{equation}
where: $h^{\left(1\right)}_{\kvec^{'}\lvec\qvec\sigma}\equiv 
c_{\kvec^{'}+\lvec+\qvec-\sigma }^{\dagger}
c_{-\kvec^{'}-\lvec-\qvec\sigma }^{\dagger}
c_{-\kvec^{'}\sigma}
c_{\kvec^{'}-\sigma}$.
By using the well known expression: $AB\simeq \left<A\right>B+A\left<B\right>-\left<A\right>\left<B\right>$, we have obtained:
\begin{equation}
\label{A2}
h^{\left(1\right)}_{\kvec^{'}\lvec\qvec\sigma}\simeq 
\Delta_{\kvec^{'}\sigma}
c_{\kvec^{'}+\lvec+\qvec-\sigma }^{\dagger}
c_{-\kvec^{'}-\lvec-\qvec\sigma }^{\dagger}
+
\Delta^{\star}_{\kvec^{'}+\lvec+\qvec\sigma}
c_{-\kvec^{'}\sigma}c_{\kvec^{'}-\sigma}
-
\Delta_{\kvec^{'}\sigma}\Delta^{\star}_{\kvec^{'}+\lvec+\qvec\sigma}.
\end{equation}
The symbol $\Delta_{\kvec\sigma}$ is given by: $\Delta_{\kvec\sigma}\equiv\left<c_{-\kvec\sigma}c_{\kvec-\sigma}\right>$.
In the next step, we have substituted \eq{A2} into \eq{A1}. The Hamiltonian can be rewritten as:
\begin{eqnarray}
\label{A3}
H_{int}&\simeq&
\frac{U}{24N^{3}}\sum^{\omega_{0}}_{\kvec\kvec^{'}\qvec\lvec\sigma}\Delta^{\star}_{\kvec^{'}+\lvec+\qvec\sigma}c_{\kvec^{'}-\sigma}
h^{\left(2\right)}_{\kvec\lvec\sigma}c_{-\kvec^{'}\sigma}
-
\frac{U}{24N^{3}}\sum^{\omega_{0}}_{\kvec\kvec^{'}\qvec\lvec\sigma}\Delta_{\kvec^{'}\sigma}c^{\dagger}_{\kvec^{'}+\lvec+\qvec-\sigma}
h^{\left(2\right)}_{\kvec\lvec\sigma}c^{\dagger}_{-\kvec^{'}-\lvec-\qvec\sigma}\\ \nonumber
&+&
\frac{U}{24N^{3}}\sum^{\omega_{0}}_{\kvec\kvec^{'}\qvec\lvec\sigma}\Delta^{\star}_{\kvec^{'}+\lvec+\qvec\sigma}\Delta_{\kvec^{'}\sigma}
h^{\left(2\right)}_{\kvec\lvec\sigma},
\end{eqnarray}
where: 
$h^{\left(2\right)}_{\kvec\lvec\sigma}\equiv c_{-\kvec+\lvec-\sigma }^{\dagger}c_{\kvec-\lvec\sigma }^{\dagger}
c_{\kvec\sigma}c_{-\kvec-\sigma}$. We notice that the linear terms with reference to the anomalous thermal average ($\Delta_{\kvec\sigma}$) can exist only on this stage of the presented approximation.  

The operator \eq{A3} has still very complicated form (the first and second terms comprise the six fermion operators). In order to solve the problem we have simplified the Hamiltonian \eq{A3} again:
\begin{equation}
\label{A4}
h^{\left(2\right)}_{\kvec\lvec\sigma}\simeq 
\Delta_{-\kvec\sigma}c_{-\kvec+\lvec-\sigma }^{\dagger}c_{\kvec-\lvec\sigma }^{\dagger}
+
\Delta^{\star}_{-\kvec+\lvec\sigma}c_{\kvec\sigma}c_{-\kvec-\sigma}
-
\Delta_{-\kvec\sigma}\Delta^{\star}_{-\kvec+\lvec\sigma}.
\end{equation}

Joining the expressions \eq{A3} and \eq{A4}, we have obtained the sum of the terms, which are proportional to $\Delta^{2}_{\kvec\sigma}$, $\Delta^{3}_{\kvec\sigma}$ or $\Delta^{4}_{\kvec\sigma}$. Next, long in the form but straightforward calculations (in the presented mean-field scheme)
give: 
\begin{widetext}
\begin{eqnarray}
\label{A5}
H_{int}&\equiv&
\frac{U}{24N^{3}}\sum^{\omega_{0}}_{\kvec\kvec^{'}\qvec\lvec\sigma}
\Delta_{\kvec\sigma}\Delta_{\kvec^{'}\sigma}\Delta^{\star}_{\kvec^{'}+\lvec+\qvec\sigma}\Delta^{\star}_{\kvec+\lvec\sigma}\\ \nonumber
&+&
\frac{U}{24N^{3}}\sum^{\omega_{0}}_{\kvec\qvec\lvec\sigma}
\Delta_{\kvec\sigma}\Delta^{\star}_{\kvec+\lvec+\qvec\sigma}
\left(c_{-\kvec-\lvec\sigma}^{\dagger}c_{-\kvec-\lvec\sigma}
+c_{-\kvec-\qvec\sigma}c_{-\kvec-\qvec\sigma}^{\dagger}\right)\\ \nonumber
&-&\frac{U}{12N^{3}}\sum^{\omega_{0}}_{\kvec\kvec^{'}\qvec\lvec\sigma}
\Delta_{\kvec\sigma}\Delta^{\star}_{\kvec^{'}+\lvec\sigma}
c_{-\kvec^{'}\sigma}c_{\kvec^{'}-\sigma}
c_{\kvec+\lvec+\qvec-\sigma}^{\dagger}c_{-\kvec-\lvec-\qvec\sigma}^{\dagger}.
\end{eqnarray}
\end{widetext}

On the basis of the expression \eq{A5}, we see that in the obtained Hamlitonian exist only the terms, which are proportional to $\Delta^{2}_{\kvec\sigma}$ or $\Delta^{4}_{\kvec\sigma}$ (all operators with $\Delta^{3}_{\kvec\sigma}$ have reduced each other mutually).
We notice that the first term in Eq. \eq{A5} can be neglected because it does not include the operators. In the case when we separate the momentums in the remained expression, the second term also can be neglected. The third term gives the interaction part of the Hamiltonian \eq{r9(3.1)} after using the mean-field approximation.

It is clear that the used scheme of the simplification of the Hamiltoniam \eq{A1} is relatively simple. However, the presented approach lets us to analyse the physics of the considered system on the quantitative level. Thus, the agreement between obtained theoretical results and the experimental data will be an essential argument that the presented method is correct. The properties of the operator \eq{A1} will be also studied diagramatically in the future. It will be done in order to improve eventually the theoretical description.      

\section{The van Hove and generalized mean-field thermodynamic potential}

In the first step, we will calculate the thermodynamic potential in the framework of the BCS van Hove scenario. Next, we will generalize the results for the case $U\neq 0$.

For $U=0$ the thermodynamic potential is given by:
\begin{equation}
\label{B1}
\Omega\left(V^{'}\right)=-\frac{1}{\beta}
\ln\left[Z\left(V^{'}\right)\right],
\end{equation}
where the grand partition function has the form:
\begin{equation}
\label{B2}
Z\left(V^{'}\right)\equiv
{\rm Tr}\left[e^{-\beta\left(H_{A}-\frac{V^{'}}{2}H_{B}\right)}\right].
\end{equation}
The symbols $ H_{A}$ and $H_{B}$ denote the following operators:
\begin{equation}
\label{B3}
H_{A}\equiv\sum_{\kvec\sigma}\overline\varepsilon _{\kvec}
c^{\dagger}_{\kvec\sigma}c_{\kvec\sigma},
\end{equation}
and
\begin{equation}
\label{B4}
H_{B}\equiv\frac{1}{N}\sum^{\omega_{0}}_{\kvec\kvec^{'}\sigma}
c^{\dagger}_{\kvec-\sigma}c^{\dagger}_{-\kvec\sigma}
c_{-\kvec^{'}\sigma}c_{\kvec^{'}-\sigma}.
\end{equation}
The thermodynamic potential is readily found from the expression:
\begin{eqnarray}
\label{B5}
& &\frac{\partial \Omega\left(V^{'}\right)}{\partial V^{'}}=
-\frac{1}{\beta Z\left(V^{'}\right)}\\ \nonumber
& &
\frac{\partial}{\partial V^{'}}{\rm Tr}
\left[\sum^{+\infty}_{j=0}\frac{1}{j!}
\left(-\beta\left(H_{A}-\frac{V^{'}}{2}H_{B}\right)\right)^{j}\right]\\ \nonumber
&=&-\frac{1}{2}\left<H_{B}\right>.
\end{eqnarray}
Integrate Eq. \eq{B5} from $V^{'}=0$ to $V^{'}=V$ we obtain:
\begin{eqnarray}
\label{B6}
\Delta\Omega_{V}&\equiv&
\frac{1}{N}\left[\Omega\left(V\right)-\Omega\left(0\right)\right]\\ \nonumber
&=&-\frac{1}{2N}\int^{V}_{0}dV^{'}\left<H_{B}\right>\\ \nonumber
&\simeq& -\int^{V}_{0}dV^{'}\left(\frac{1}{V^{'}}\right)^{2}|\Delta_{V^{'}}|^{2},
\end{eqnarray}
where $\Delta_{V^{'}}\equiv V^{'}\Delta$.
The formula \eq{B6} may be rewritten as follows:
\begin{equation}
\label{B7}
\Delta\Omega_{V}=
\int^{\Delta_{V}}_{0}d\Delta_{V^{'}}\left(\Delta_{V^{'}}\right)^{2}
\frac{d\left(\frac{1}{V^{'}}\right)}{d\Delta_{V^{'}}}.
\end{equation}
After substituting Eq. \eq{r5(3.2)} (for $U=0$) into expression \eq{B7} we find:
\begin{eqnarray}
\label{B8}
\Delta\Omega_{V}&=&
\frac{\Delta^{2}_{V}}{V}
-2\int_{0}^{\omega_{0}}d\varepsilon\rho\left(\varepsilon\right)
\left(E-\varepsilon\right)\\ \nonumber
&+&\frac{4}{\beta}\int_{0}^{\omega_{0}}d\varepsilon\rho\left(\varepsilon\right)
\ln\left(1+e^{-\beta\varepsilon}\right)\\ \nonumber
&-&\frac{4}{\beta}\int_{0}^{\omega_{0}}d\varepsilon\rho\left(\varepsilon\right)
\ln\left(1+e^{-\beta E}\right).
\end{eqnarray}
The first integral in Eq. \eq{B8} is given by:
\begin{eqnarray}
\label{B9}
I_{1}\left(\Delta_{V}\right)&\equiv& 
-2\int_{0}^{\omega_{0}}d\varepsilon\rho\left(\varepsilon\right)
\left(E-\varepsilon\right)\\ \nonumber
&=&b_{1}\Delta^{2}_{V}\sum^{3}_{j=1}f_{j}\left(\Delta_{V}\right),
\end{eqnarray}
where:
\begin{widetext}
\begin{equation}
\label{B10}
f_{1}\left(\Delta_{V}\right)\equiv
\left(\frac{\omega_{0}}{\Delta_{V}}\right) 
F_{3,2}\left[\frac{1}{2},\frac{1}{2},\frac{1}{2};\frac{3}{2},
\frac{3}{2};-\left(\frac{\omega_{0}}{\Delta_{V}}\right)^{2}\right],
\end{equation}
\begin{equation}
\label{B11}
f_{2}\left(\Delta_{V}\right)\equiv
\left[\frac{1}{2}-\ln\left(\omega_{0}\right)\right]
\left[
\left(\frac{\omega_{0}}{\Delta_{V}}\right)^{2}\left[\sqrt{1+\left(\frac{\Delta_{V}}{\omega_{0}}\right)^{2}}-1\right]
+\rm{arcsinh}\left(\frac{\omega_{0}}{\Delta_{V}}\right)
\right],
\end{equation}
and
\begin{equation}
\label{B12}
f_{3}\left(\Delta_{V}\right)\equiv
\ln\left(b_{2}\right)
\left[
\left(\frac{\omega_{0}}{\Delta_{V}}\right)^{2}\left[\sqrt{1+\left(\frac{\Delta_{V}}{\omega_{0}}\right)^{2}}-1\right]
+
\ln\left[\left(\frac{\omega_{0}}{\Delta_{V}}\right)\left[\sqrt{1+\left(\frac{\Delta_{V}}{\omega_{0}}\right)^{2}}+1\right]\right]
\right].
\end{equation}
\end{widetext}
The symbol $F_{p,q}\left(a;b;z\right)$ denotes the generalized hypergeometric function. 

The second integral in Eq. \eq{B8} describes the first temperature-dependent correction to the thermodynamic potential in the normal state:
\begin{equation}
\label{B13}
I_{2}\left(T\right)\equiv\frac{4}{\beta}
\int_{0}^{\omega_{0}}d\varepsilon\rho\left(\varepsilon\right)
\ln\left(1+e^{-\beta\varepsilon}\right).
\end{equation}
The expression \eq{B13} can be rewritten by using the partial integration method. Since $\omega_{0}>>k_{B}T$, the obtained integral may be extended to infinity. In this way, we can find:
\begin{eqnarray}
\label{B14}
& &I_{2}\left(T\right)=-b_{1}\left[\kappa+\ln\left(b_{2}\beta\right)\right]
\frac{\pi^{2}}{3\beta^{2}}
\\ \nonumber
& &
+b_{1}\omega_{0}\left[\ln\left(\omega_{0}\right)-1\right]
\ln\left(1+e^{-\beta\omega_{0}}\right)\frac{4}{\beta}.
\end{eqnarray}

The number $\kappa\simeq 0.45403$ is defined by:
\begin{eqnarray}
\label{B15}
\kappa&\equiv&1+\ln\left(2\right)+\gamma\\ \nonumber
&+&\frac{3}{\pi^{2}}
\left[\left(\frac{\partial \gamma_{1}\left(z\right)}{\partial z}\right)_{z=1}-\left(\frac{\partial \gamma_{1}\left(z\right)}{\partial z}\right)_{z=\frac{1}{2}}\right],
\end{eqnarray}
where the symbol $\gamma_{n}\left(z\right)$ is the generalized Stieltjes constant. 

We also notice that, by using the equations \eq{B7} and \eq{r13(3.2)} it is possible to derive explicit expression for $\Delta\Omega_{V}$ close to the transition temperature:
\begin{equation}
\label{B16}
\Delta\Omega_{V}=-\frac{b_{1}}{2 p_{2}\left(T\right)}
\left[p_{1}\left(T\right)-\frac{1}{b_{1}V}\right]^{2}.
\end{equation}

\begin{figure}[t]%
\includegraphics*[scale=0.15]{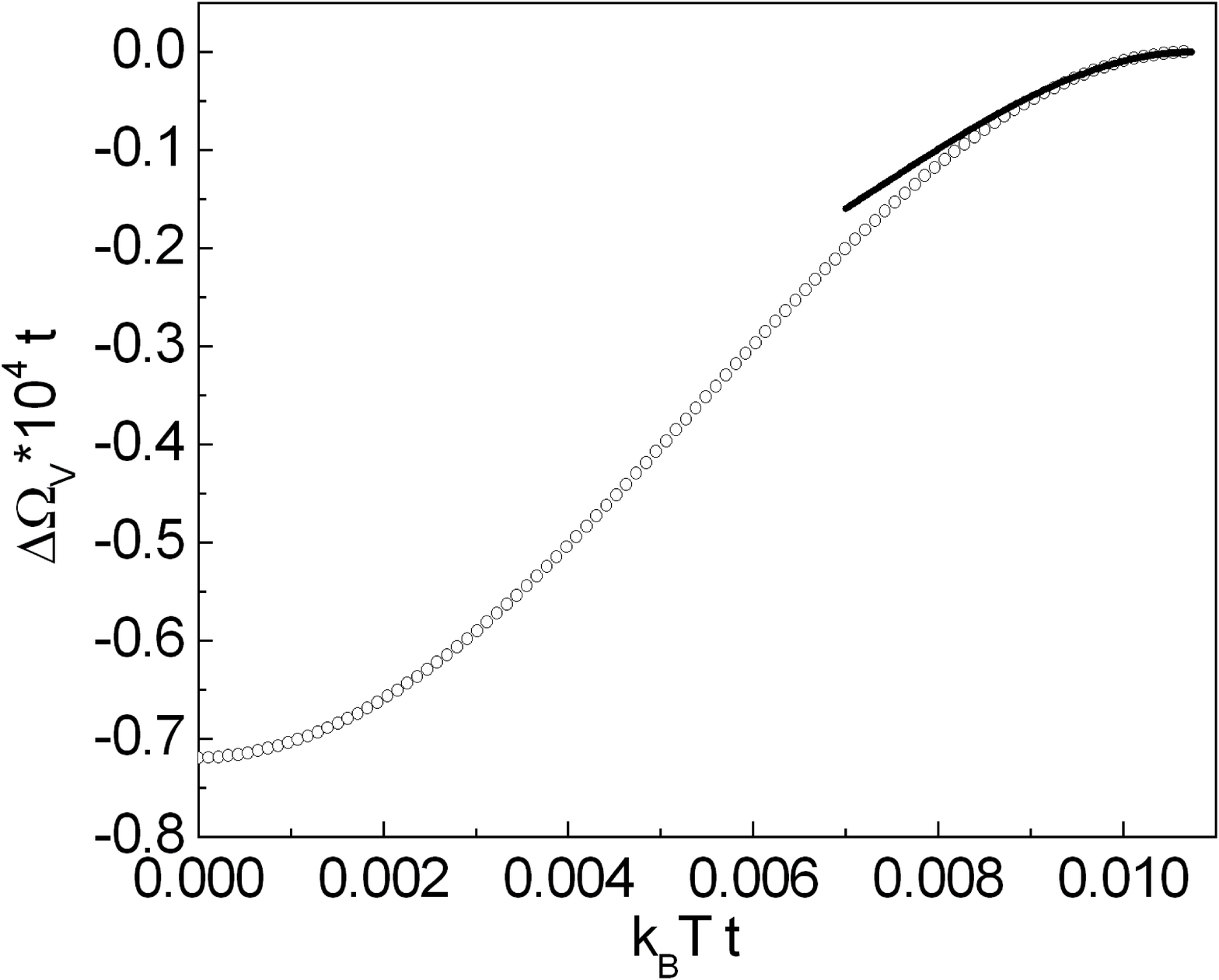}
\caption{The dependence of $\Delta\Omega_{V}$ on the temperature. We assume $V=1t$ and $\omega_{0}=0.3t$. 
The empty circles are obtained from Eq. \eq{B8} with help of Eqs. \eq{B9} and \eq{B14}. 
Solid line represents the calculation of $\Delta\Omega_{V}$ using the formula \eq{B16}.}
\label{f12} 
\end{figure}

The dependence of $\Delta\Omega_{V}$ on the temperature is shown in Fig. \ref{f12}.
%

The calculation of the thermodynamic potential for $U\neq 0$ is more difficult problem. In this case:
\begin{equation}
\label{B17}
\Omega\left(V^{'},U^{'}\right)=-\frac{1}{\beta}
\ln\left[Z\left(V^{'},U^{'}\right)\right],
\end{equation}
where the grand partition function has the form:
\begin{equation}
\label{B18}
Z\left(V^{'},U^{'}\right)\equiv
{\rm Tr}\left[e^{-\beta\left(H_{A}-\frac{V^{'}}{2}H_{B}-\frac{U^{'}}{24}
H_{C}\right)}\right].
\end{equation}
In Eq. \eq{B18} the Hamiltonian $H_{C}$ is given by:
\begin{widetext}
\begin{equation}
\label{B19}
H_{C}\equiv\frac{1}{N^{3}}\sum_{\kvec\kvec^{'}\qvec\lvec\sigma}
c_{\kvec-\lvec\sigma }^{\dagger}c_{\kvec\sigma}
c_{\kvec^{'}+\lvec+\qvec-\sigma }^{\dagger}c_{\kvec^{'}-\sigma}
c_{-\kvec^{'}-\lvec-\qvec\sigma }^{\dagger}c_{-\kvec^{'}\sigma}
c_{-\kvec+\lvec-\sigma }^{\dagger}c_{-\kvec-\sigma}.
\end{equation}
\end{widetext}

Now, we consider the total differential of the thermodynamic potential: 
\begin{equation}
\label{B20}
d\Omega\left(V^{'},U^{'}\right)=
\frac{\partial \Omega\left(V^{'},U^{'}\right)}{\partial V^{'}}dV^{'}+
\frac{\partial \Omega\left(V^{'},U^{'}\right)}{\partial U^{'}}dU^{'}.
\end{equation}
By using the method presented for the BCS van Hove scenario we can obtain:
\begin{equation}
\label{B21}
\frac{\partial \Omega\left(V^{'},U^{'}\right)}{\partial V^{'}}
=-\frac{1}{2}\left<H_{B}\right>\simeq-N|\Delta\left(V^{'},U^{'}\right)|^{2},
\end{equation}
and
\begin{equation}
\label{B22}
\frac{\partial \Omega\left(V^{'},U^{'}\right)}{\partial U^{'}}=
-\frac{1}{24}\left<H_{C}\right>\simeq-\frac{N}{12}|\Delta\left(V^{'},U^{'}\right)|^{4}.
\end{equation}

Finally, the general evaluation of the thermodynamic potential requires the numerical analysis of the expression:
\begin{widetext}
\begin{eqnarray}
\label{B23}
\Delta\Omega_{VU}&\equiv&
\frac{1}{N}\left[\Omega\left(V,U\right)-\Omega\left(0,0\right)\right]\\ \nonumber
&\simeq&
-\int_{\left(0,0\right)}^{\left(V,U\right)}|\Delta\left(V^{'},U^{'}\right)|^{2}dV^{'}+
\frac{1}{12}|\Delta\left(V^{'},U^{'}\right)|^{4}dU^{'}
=
-\int_{0}^{1}dx
\left[V|\Delta\left(Vx,Ux\right)|^{2}+\frac{U}{12}|\Delta\left(Vx,Ux\right)|^{4}\right].
\end{eqnarray}
\end{widetext}

\newpage

\section{Experimental values of $T_{C}$ and the low-temperature superconducting gap}

In the Appendix we provide the list of the thermodynamic parameters values of high-$T_{C}$ superconductors which have been obtained experimentally. In particular, we have collected the data for $T_{C}$ and the energy gap $\Delta_{tot}^{\left(0\right)}$. We have also determined the doping level or stoichiometry of the materials and the values of $R_{1}$ parameter. 

\begin{table}[h]
\caption{\label{t2} The experimental data for ${\rm YBa_{2}Cu_{3}O_{7-y}}$ (YBCO).}
\begin{ruledtabular}
\begin{tabular}{ccccc}
Type&$T_{C}$ (K)&$\Delta_{tot}^{\left(0\right)}$ (meV)&$R_{1}$&Ref.\\
\hline
$p=0.079$ & 44   & 66 & 34.81 & \cite{Sutherland}\\
$p=0.111$ & 62   & 71 & 26.58 &                  \\
$p=0.166$ & 93.5 & 50 & 12.41 &                  \\
$p=0.184$ & 89   & 37 & 9.65  &                  \\
\hline
$p=0.106$ & 60   & $58\pm 8.8$  & $22.37\pm 3.39$ &  \cite{Nakayama}\\
$p=0.137$ & 80   & $45\pm 4.9$  & $13.04\pm 1.40$ &                 \\
$p=0.175$ & 92   & $34\pm 3.1$  & $8.55 \pm 0.77$ &                 \\
\hline
$p=0.099$ & 57.4 & 56.8 & 22.96 & \cite{Kaminski}\\
$p=0.110$ & 61.8 & 54.2 & 20.36 &                \\
$p=0.140$ & 83   & 37.6 & 10.52 &                \\
$p=0.160$ & 93.2 & 33.8 & 8.42  &                \\
\hline
$p=0.080$  & 46.2 & 66.3   & 33.27 & \cite{Plate}\\
$p=0.096$  & 56.3 & 71     & 29.3  &             \\
$p=0.160$  & 93.2 & 49.9   & 12.42 &             \\
$p=0.184$  & 88.9 & 37.2   & 9.7   &             \\
\hline
$p=0.086$  &  51.6  &  12.5  &  5.62  & \cite{Morr}, \cite{Fong}\\
$p=0.096$  &  56.3  &  14    &  5.77  &                   \\
$p=0.101$  &  58    &  16.5  &  6.6   &                   \\
$p=0.171$  &  93    &  20.5  &  5.12  &                   \\
\hline
$p=0.106$  &  60    &  $\sim 20$  & 7.7 & \cite{Yeh}\\
$p=0.159$  &  92.9  &  18         & 4.5 &           \\
\hline
$p=0.149$  &  89    &  20         & 5.2 & \cite{Born}\\
\hline
$p=0.153$  &  91  &  24-32                                                              & 7.1 & \cite{Murakami}\\
$p=0.153$  &  91  &  $\sim 25$\footnote{Tunneling on electrical field etched surface.}  & 6.4 &                \\
\hline
$p=0.156$  &  92  &  $30\pm 8$  & 7.6   & \cite{Edwards1}\\
\hline
$p=0.156$  &  92  &  20         & 5     & \cite{Edwards2}\\  
\end{tabular}
\end{ruledtabular}
\end{table}

\begin{table}[h]
\caption{\label{t3} The experimental data for ${\rm NdBa_{2}Cu_{3}O_{7-y}}$ (NdBCO).}
\begin{ruledtabular}
\begin{tabular}{ccccc}
Type&$T_{C}$ (K)&$\Delta_{tot}^{\left(0\right)}$ (meV)&$R_{1}$&Ref.\\
\hline
y=0 & 95 & 30 & 7.3 & \cite{Nishiyama}\\
\end{tabular}
\end{ruledtabular}
\end{table}

\newpage

\begin{table}[h!]
\caption{\label{t4} The experimental data for ${\rm Bi_{2}Sr_{2}CaCu_{2}O_{8+y}}$ (Bi2212)}
\begin{ruledtabular}
\begin{tabular}{ccccc}
Type&$T_{C}$ (K)&$\Delta_{tot}^{\left(0\right)}$ (meV)&$R_{1}$&Ref.\\
\hline
$p=0.125$ & 83     & $44  \pm 2$   &   $12.30\pm 0.56$   &   \cite{Renner}\\
$p=0.160$ & 92.2   & $41.5\pm 2$   &   $10.45\pm 0.5 $   &                \\
$p=0.208$ & 74.3   & $34  \pm 2$   &   $10.62\pm 0.62$   &                \\
$p=0.229$ & 56     & $21  \pm 2$   &   $ 8.7 \pm 0.83$   &                \\
\hline
$p=0.131$ & 86     & $32.2\pm 1.1$    & $8.7\pm 0.3$   & \cite{Hoffmann}\\
$p=0.198$ & 81     & $28.6\pm 1.4$    & $8.2\pm 0.4$   &                \\
\hline
$p=0.189$ & $86\pm 4$     & $25\pm 1$    & $6.7\pm 0.5$   & \cite{Ponomarev}\\
\hline
$p=0.186$ & $87\pm 1$    & $26\pm 1$       & $8.2\pm 0.3$   & \cite{Oki}\\
$p=0.165$ & $92\pm 3$    & $35\pm 1$       & $8.9\pm 0.7$   &           \\
$p=0.134$ & $\sim 87$    & $\sim 45$       & $\sim 12$      &           \\
$p=0.127$ & $84\pm 4$    & $32\pm 0.5$     & $8.8\pm 0.6$   &           \\
$p=0.123$ & $82\pm 4$    & $33.5\pm 0.5$   & $9.4\pm 0.6$   &           \\
\hline
$p=0.160$  & 92.5    & 32.5   &   8.15   & \cite{Krasnov}\\
$p=0.181$ & 89      & 25.8   &   6.73   &               \\
\hline
$p=0.106$ & 70     & 38           & 12.6           & \cite{Gupta}\\
$p=0.191$ & 85     & $30\pm 2$    & $8.2\pm 0.5$   &             \\
\hline
$p=0.103$ & 67     & 39.8  & 13.79 & \cite{Kanigel}\\
$p=0.122$ & 80     & 35.9  & 10.42 &               \\
\hline
$p=0.086$ & 50.9     & 64.5  & 29.39 & \cite{Campuzano}, \cite{Tanaka}.\\
$p=0.089$ & 54.2     & 61.2  & 26.21 &                                 \\
$p=0.110$ & 73.2     & 47.8  & 15.16 &                                 \\
$p=0.115$ & 76.8     & 50.1  & 15.14 &                                 \\
$p=0.121$ & 80.6     & 46.1  & 13.27 &                                 \\
$p=0.133$ & 86.7     & 43.5  & 11.65 &                                 \\
$p=0.161$ & 92.2     & 37.5  & 9.43  &                                 \\
$p=0.186$ & 87       & 31    & 8.28  &                                 \\
$p=0.193$ & 83.8     & 36.6  & 10.14 &                                 \\
$p=0.201$ & 79.3     & 25.8  & 7.56  &                                 \\
$p=0.205$ & 76.6     & 34    & 10.32 &                                 \\
$p=0.215$ & 69       & 27.2  & 9.13  &                                 \\
\hline
$p=0.100$ & 63     & 40    & 14.7 &               \cite{Nakaono}\\
$p=0.130$ & 85     & 33    & 9.0  &                             \\
$p=0.190$ & 85     & 26    & 7.1  &                             \\
\hline
$p=0.100$ & 60   & $36 \pm 2$  & 13.9 &               \cite{Oda}\\
$p=0.140$ & 82   & $34 \pm 2$  & 9.6  &                         \\
$p=0.160$ & 88   & $32 \pm 2$  & 8.4  &                         \\
$p=0.210$ & 81   & $27 \pm 2$  & 7.7  &                         \\
\hline
$p=0.110$ &   65    & 62          & 22.1 &               \cite{McElroy}\\
$p=0.130$ &   75    & $48\pm 1$   & 14.9 &                             \\
$p=0.150$ &   79    & $43\pm 1$   & 12.6 &                             \\
$p=0.180$ &   89    & $36\pm 1$   & 9.4  &                             \\
$p=0.190$ &   89    & $33\pm 1$   & 8.6  &                             \\
\hline
$p=0.110$ & 67     & $55\pm 15$  & 19.1 &               \cite{Matsuda}\\
$p=0.130$ & 85     & $45\pm 12$  & 12.3 &                             \\
$p=0.160$ & 89     & $40\pm 10$  & 10.4 &                             \\
$p=0.180$ & 89     & $35\pm 7$   & 9.1  &                             \\
$p=0.220$ & 64     & $22\pm 5$   & 8    &                             \\
\hline
$p=0.120$ & 78     & 50.2  &   14.9 &               \cite{Hoffman}\\
$p=0.160$ & 92     & 43.7  &   11   &                             \\
$p=0.190$ & 85     & 36.7  &   10   &                             \\
\hline
$p=0.120$ & 80     & $42\pm 2$  & 12.2 &               \cite{Howald}\\
\hline
$p=0.120$ & 81     & 40  & 11.5 &               \cite{Murakami1}\\
\end{tabular}
\end{ruledtabular}
\end{table}

\newpage

\begin{table}[h]
\caption{\label{t5} The experimental data for ${\rm Bi_{2}Sr_{2}Ca_{2}Cu_{3}O_{10+y}}$ (Bi2223).}
\begin{ruledtabular}
\begin{tabular}{ccccc}
Type&$T_{C}$ (K)&$\Delta_{tot}^{\left(0\right)}$ (meV)&$R_{1}$&Ref.\\
\hline
OP & $110\pm 5$ & $36 \pm 1.6$ & $7.6 \pm 0.5$  &\cite{Ponomarev}\\
\hline
UD & 109 & $60\pm 3$ & 12.8 & \cite{Kugler3}\\
OP & 111 & $45\pm 7$ & 9.4  &               \\
\hline
OP &  109       & $\sim 37$ & $\sim 7.88$    &\cite{Masui}
\end{tabular}
\end{ruledtabular}
\end{table}

\begin{table}[h]
\caption{\label{t6} The experimental data for ${\rm Pr_{2-x}Ce_{x}CuO_{4-y}}$ (PCCO).}
\begin{ruledtabular}
\begin{tabular}{ccccc}
Type&$T_{C}$ (K)&$\Delta_{tot}^{\left(0\right)}$ (meV)&$R_{1}$&Ref.\\
\hline
x=0.13 & 12.2 & $6.3\pm 0.6$ & $11.99\pm 1.2$  & \cite{Biswas}\\
x=0.15 & 21.6 & $4.6\pm 0.3$ & $4.95\pm 0.36$  &              \\
x=0.17 & 11.8 & $1.8\pm 0.2$ & $3.46\pm 0.45$  &              \\
\hline
x=0.15 & 21 & (4.3-5.4) & (4.7-6)   & \cite{Zimmers1}\\
x=0.17 & 15 & (3-3.6) & (4.7-5.6) &                \\
\hline
x=0.15     &  $19\pm 1$  &  3.25  &  $4  \pm 0.4$     & \cite{Dagan}\\
x=0.16     &  $16\pm 1$  &  2.6   &  $3.8\pm 0.5$     &             \\
x=0.17     &  $13\pm 1$  &  1.3   &  $2.3\pm 1.3$     &             \\
x=0.18     &  $11\pm 1$  &  1.0   &  $2.1\pm 1.5$     &             \\
x=0.19     &  $8\pm 0.4$ &  0.9   &  $2.6\pm 0.9$     &             \\
\hline
x=0.15 & 20 & $4.3$ & 5 & \cite{Homes}\\
\hline
x=0.13 & 10 & 6.5  & 15.09 & \cite{Fournier}\\
x=0.15 & 23 & 4.4  & 4.4   &                \\
x=0.17 & 13 & 1.75 & 3.12  &                \\
\end{tabular}
\end{ruledtabular}
\end{table}

%
 
%

\begin{thebibliography}{[1]}
%
\bibitem{Bednorz}
(a) J.G. Bednorz, K.A. Muller, Z. Phys. B {\bf 64}, 189 (1986);\\
(b) J.G. Bednorz, K.A. Muller, Rev. Mod. Phys. {\bf 60}, 585 (1988).
\bibitem{Dagotto}
E. Dagotto, Rev. Mod. Phys. {\bf 66}, 763 (1994).
\bibitem{Hubbard}
J. Hubbard, Proc. R. Soc. London, Ser. A {\bf 276}, 238 (1963).
\bibitem{Emery1}
(a) V.J. Emery, Phys. Rev. Lett. {\bf 58}, 2794 (1987);\\
(b) P.B. Littlewood, C.M. Varma, E. Abrahams, Phys. Rev. Lett. {\bf 60}, 379 (1987).
\bibitem{Anderson}
P.W. Anderson, Science {\bf 235}, 1196 (1987).
\bibitem{Lee1}
P.A. Lee, N. Nagaosa, C.-G. Wen, Rev. Mod. Phys. {\bf 78}, 17 (2006).
\bibitem{Millis}
(a) A.J. Millis, H. Monien, D. Pines, Phys. Rev. B {\bf 42}, 167 (1990);\\
(b) P. Monthoux, D. Pines, Phys. Rev. Lett. {\bf 69}, 961 (1992);\\ 
(c) R.J. Radtke, K. Levin, H.-B. Schuttler, M.R. Norman, Phys. Rev. B {\bf 48}, 15957 (1993).
\bibitem{Spalek}
K.A. Chao, J. Spa{\l}ek, A.M. Ole{\'s}, J. Phys.: Solid State C {\bf 10}, L271 (1977).
\bibitem{Kulic}
(a) J.H. Kim, Z. Tesanovic, Phys. Rev. Lett. {\bf 71}, 4218 (1993);\\
(b) M.L. Kulic, Lectures on the physics of highly correlated electron systems VIII: 8th Training Course in the Physics of
Correlated Electron Systems and High-$T_{C}$ Superconductors. AIP Conference Proceedings {\bf 715}, 75 (2004);\\
(c) M.L. Kulic, Journal of Superconductivity and Novel Magnetism {\bf 19}, 213 (2006).
\bibitem{Hybertsen}
M.S. Hybertsen, E.B. Stechel, M. Schluter, D.R. Jennison, Phys. Rev. B {\bf 41}, 11068 (1990).
\bibitem{Imada}
(a) M. Imada, Y. Hatsugai, J. Phys. Soc. Jpn. {\bf 58}, 3752 (1989);\\
(b) M. Imada, J. Phys. Soc. Jpn. {\bf 60}, 2740 (1991);\\
(c) D.J. Scalapino, S.R. White, S.C. Zhang, Phys. Rev. Lett. {\bf 68}, 2830 (1992);\\
(d) J.E. Hirsch, in: Proceedings of the International Conference on Strongly Correlated Electron Systems, San Diego, August 1993.
\bibitem{Pryadko}
L. Pryadko, S. Kivelson, O. Zachar, Phys. Rev. Lett. {\bf 92}, 067002 (2004).
\bibitem{Franck}
J.P. Franck, in: Physical Properties of High Temperature Superconductors,
edited by D.M. Ginsberg (World Scientific, Singapore, 1994), Vol. IV, p. 189.
\bibitem{Kulic2}
M.L. Kulic, Phys. Rep. {\bf 338}, 1 (2000).
\bibitem{Vedeneev}
(a) S.I. Vedeneev, A.G.M. Jansen, A.A. Tsvetkov, P. Wyder, Phys. Rev. B {\bf 51}, 16380 (1995);\\ 
(b) C.C. Tsuei, J.R. Kirtley, M. Rupp, A. Gupta, J.Z. Sun, T. Shaw, M.B. Ketchen, C. Wang, 
    Z.F. Ren, J. H. Wang, M. Bhushan Science {\bf 27}, 329 (1996);\\
(c) C.C. Tsuei, J.R. Kirtley, Z.F. Ren, J.H. Wang, H. Raffy, Z.Z. Li, Nature {\bf 387}, 481 (1998).
\bibitem{Hofer}
(a) J. Hofer, K. Conder, T. Sasagawa, Guo-meng Zhao, M. Willemin, H. Keller, K. Kishio, 
    Phys. Rev. Lett. {\bf 84}, 4192 (2000);\\ 
(b) T. Schneider, Phys. Stat. Sol. (b) {\bf 242}, 58 (2005).
\bibitem{Damascelli}
A. Damascelli, Z. Hussain, Z.-X. Shen, Rev. Mod. Phys. {\bf 75}, 473 (2003).
\bibitem{Cuk}
T. Cuk, D.H. Lu, X.J. Zhou, Z.-X. Shen, T.P. Deveraux, N. Nagaosa, Phys. Stat. Sol. (b) {\bf 242}, 11 (2005).
\bibitem{Gweon}
G.-H. Gweon, T. Sasagawa, S.Y. Zhou, J. Graf, H. Takagi, D.-H. Lee, A. Lanzara, Nature {\bf 430}, 187 (2004).
\bibitem{Heid}
R. Heid, R. Zeyher, D. Manske, K.-P. Bohnen, Phys. Rev. B {\bf 80}, 024507 (2009).
\bibitem{Bohnen}
K.-P. Bohnen, R. Heid, M. Krauss, Europhys. Lett. {\bf 64}, 104 (2003).
\bibitem{VanHove}
L. van Hove, Phys. Rev. {\bf 89}, 1189 (1953).
\bibitem{Markiewicz}
R.S. Markiewicz, J. Phys. Chem. Sol. {\bf 58}, 1179 (1997).
\bibitem{Frohlich}
(a) H. Fr{\"o}hlich, Phys. Rev. {\bf 79}, 845 (1950);\\
(b) H. Fr{\"o}hlich, Proc. R. Soc. A {\bf 223}, 296 (1954).
\bibitem{Hirsch}
J.E. Hirsch, Phys. Rev. Lett. {\bf 87}, 206402 (2001).
\bibitem{Hirsch1}
J.E. Hirsch, Phys. Rev. B {\bf 66}, 064507 (2002).
\bibitem{Marsiglio}
F. Marsiglio, R. Teshima, J.E. Hirsch, Phys. Rev. B {\bf 68}, 224507 (2003).
\bibitem{Lang}
I.G. Lang, Y.A. Firsov, Zh. Eksp. Teor. Fiz. {\bf 43}, 923 (1962).
\bibitem{Eliashberg}
For discussion of the Eliashberg equations [originally formulated by G.M. Eliashberg, Soviet.
Phys. JETP {\bf 11}, 696 (1960)] we refer to: \ \\
(a) P.B. Allen, B. Mitrovi{\'c}, in: Solid State
Physics: Advances in Research and Applications,
edited by H. Ehrenreich, F. Seitz, D. Turnbull,
 (Academic, New York, 1982), Vol 37, p. 1;   \ \\
(b) J.P. Carbotte, Rev. Mod. Phys. {\bf 62}, 1027 (1990); \ \\
(c) J.P. Carbotte, F. Marsiglio, in: The Physics of Superconductors, edited by K.H. Bennemann, J.B. Ketterson, (Springer, Berlin, 2003), Vol 1, p. 223.
\bibitem{Bardeen}
(a) J. Bardeen, L.N. Cooper, J.R. Schrieffer, Phys. Rev. {\bf 106}, 162 (1957);\\
(b) J. Bardeen, L.N. Cooper, J.R. Schrieffer, Phys. Rev. {\bf 108}, 1175 (1957).
\bibitem{Xu}
J.H. Xu, T.J. Watson-Yang, J. Yu, A.J. Freeman, Phys. Lett. {\bf 120A}, 489 (1987).
\bibitem{Nunner}
T.S. Nunner, J. Schmalian, K.H. Bennemann, Phys. Rev. B {\bf 59}, 8859 (1999).
\bibitem{Andersen}
O.K. Andersen, A.I. Liechtenstein, O. Jepsen, F.
Paulsen, J. Phys. Chem. Solids {\bf 56}, 1573 (1995).
\bibitem{Lin}
J. Lin, A.J. Millis, Phys. Rev. B {\bf 72}, 214506 (2005).
\bibitem{Hauge}
J.P. Hauge, AIP Conference Proceedings, {\bf 846(1)}, 255 (2006).
\bibitem{Szczesniak1} 
(a) R. Szcz{\c{e}}{\'s}niak, M. Mierzejewski,
J. Zieli{\'n}ski, P. Entel, Solid State Commun. {\bf 117},
369 (2001);\\
(b) R. Szcz{\c{e}}{\'s}niak, S. Grabi{\'n}ski, Acta Phys. Polonica A {\bf 102}, 401 (2002).
\bibitem{Szczesniak3} 
R. Szcz{\c{e}}{\'s}niak, Solid State Commun. {\bf 138},
347 (2006).
\bibitem{Czerwonko}
(a) J. Czerwonko, J. Phys. Element. Part. At. Nucl. (Dubna) {\bf 31}, 145 (2000);\\
(b) J. Czerwonko, Acta Phys. Polonica B {\bf 29}, 3885 (1998).
\bibitem{Goicochea}
A.G. Goicochea, Phys. Rev. B {\bf 49}, 6864 (1994).
\bibitem{Sarkar}
(a) S. Sarkar, S. Basu, A.N. Das, Phys. Rev. B {\bf 52}, 12545 (1995);\\
(b) S. Sarkar, S. Basu, A.N. Das, Phys. Rev. B {\bf 51}, 12854 (1995);\\
(c) S. Sarkar, A.N. Das, Phys. Rev. B {\bf 54}, 14974 (1996).
\bibitem{Mamedov}
T.A. Mamedov, M. de Llano, Phys. Lett. A {\bf 257}, 201 (1999).
\bibitem{Renner}
(a) Ch. Renner, B. Revaz, J.-Y. Genoud, K. Kadowaki, O. Fischer, Phys. Rev. Lett. {\bf 80}, 149 (1998);\\
(b) Ch. Renner, B. Revaz, K. Kadowaki, I. Maggio-Aprile, O. Fischer, Phys. Rev. Lett. {\bf 80}, 3606 (1998).
\bibitem{Wang}
Y. Wang, L. Li, N.P. Ong, Phys. Rev. B {\bf 73}, 024510 (2006).
\bibitem{Liang}
R. Liang, D.A. Bonn, W.N. Hardy, Phys. Rev. B {\bf 73}, 180505(R) (2006).
\bibitem{Presland}
M.R. Presland, J.L. Tallon, R.G. Buckley, R.S. Liu, N.E. Flower, Physica C {\bf 176}, 95 (1991).
\bibitem{Ong}
N.P. Ong, Y. Wang, S. Ono, Y. Ando, S. Uchida, Ann. Phys. {\bf 13}, 9 (2004). 
\bibitem{Rullier}
F. Rullier-Albenque, R. Tourbot, H. Alloul, P. Lejay, D. Colson, A. Forget, Phys. Rev. Lett. {\bf 96}, 067002 (2006).
\bibitem{Xu1}
Z.A. Xu, J.Q. Shen, S.R. Zhao, Y.J. Zhang, C.K. Ong, Phys. Rev. B {\bf 72}, 144527 (2005).
\bibitem{Li2}
P. Li, S. Mandal, R.C. Budhani, R.L. Greene, Phys. Rev. B {\bf 75}, 184509 (2007).
\bibitem{Johannsen}
N. Johannsen, Th. Wolf, A.V. Sologubenko, T. Lorenz, A. Freimuth, J.A. Mydosh, Phys. Rev. B {\bf 76}, 020512(R) (2007).
\bibitem{Tohayama}
T. Tohayama, S. Maekawa, Supercond. Sci. Technol. {\bf 13}, R17 (2000).
\bibitem{Tohayama1}
T. Tohayama, S. Maekawa, Phys. Rev. B {\bf 67}, 092509 (2003).
\bibitem{Kim1}
C. Kim, P.J. White, Z.-X. Shen, T. Tohyama, Y. Shibata, S. Maekawa, B.O. Wells, Y.J. Kim, R.J. Birgeneau, M.A. Kastner, 
Phys. Rev. Lett. {\bf 80}, 4245 (1998).
\bibitem{Kulic1}
M.L. Kulic, O.V. Dolgov, Phys. Rev. B {\bf 76}, 132511 (2007). 
\bibitem{Gonnelli}
R.S. Gonnelli, G.A. Ummarino, V.A. Stepanov, Physica C {\bf 275}, 162 (1997). 
\bibitem{Wang1}
Y. Wang, L. Li, M.J. Naughton, G.D. Gu, S. Uchida, N.P. Ong, Phys. Rev. Lett. {\bf 95}, 247002 (2005).
\bibitem{Zimmers}
A. Zimmers, L. Shi, D.C. Schmadel, W.M. Fisher, R.L. Greene, H.D. Drew, M. Houseknecht, G. Acbas,
M.-H. Kim, M.-H. Yang, J. Cerne, J. Lin, A. Millis, Phys. Rev. B {\bf 76}, 064515 (2007).
\bibitem{Hackl}
A. Hackl, S. Sachdev, Phys. Rev. B {\bf 79}, 235124 (2009).
\bibitem{Khlopkin}
M.N. Khlopkin, G.Kh. Panova, A.A. Shikov, N.A. Chernoplekov, Phys. Solid State {\bf 41}, 1050 (1999).
\bibitem{Balci}
H. Balci, V.N. Smolyaninova, P. Fournier, A. Biswas, R.L. Greene, Phys. Rev. B {\bf 66}, 174510 (2002).
\bibitem{Li}
P. Li, R.L. Greene, Phys. Rev. B {\bf 76}, 174512 (2007).
\bibitem{Sutherland}
M. Sutherland, D.G. Hawthorn, R.W. Hill, F. Ronning, S. Wakimoto,
H. Zhang, C. Proust, E. Boaknin, C. Lupien, L. Taillefer, 
R. Liang, D.A. Bonn, W.N. Hardy, R. Gagnon, N.E. Hussey, 
T. Kimura, M. Nohara, H. Takagi, Phys. Rev. B {\bf 67}, 174520 (2003).
\bibitem{Nakayama}
K. Nakayama, T. Sato, K. Terashima, T. Arakane, T. Takahashi, M. Kubota, K. Ono, 
T. Nishizaki, Y. Takahashi, N. Kobayashi, Phys. Rev. B {\bf 79}, 140503(R) (2009).
\bibitem{Kaminski}
A. Kaminski, S. Rosenkranz, H.M. Fretwell, J. Mesot, M. Randeria, J.C. Campuzano, M.R. Norman, Z.Z. Li, H. Raffy, T. Sato, T. Takahashi, K. Kadowaki, Phys. Rev. B {\bf 69}, 212509 (2004).
\bibitem{Plate}
M. Plate, J.D.F. Mottershead, I.S. Elfimov, D.C. Peets, Ruixing Liang, D.A. Bonn, W.N. Hardy, S. Chiuzbaian, M. Falub, M. Shi, L. Patthey, A. Damascelli, Phys. Rev. Lett. {\bf 95}, 077001 (2005).
\bibitem{Morr}
D.K. Morr, D. Pines, Phys. Rev. Lett. {\bf 81}, 1086 (1998).
\bibitem{Fong}
H.F. Fong, B. Keimer, D.L. Milius, I.A. Aksay, Phys. Rev. Lett. {\bf 78}, 713 (1997). 
\bibitem{Yeh}
N.C. Yeh, C.T. Chen, G. Hammer, J. Mannhart, A. Schmehl, C.W. Schneider, R.R. Schulz, S. Tajima, 
K. Yoshida, D. Garrigus, M. Strasik, Phys. Rev. Lett. {\bf 87}, 087003 (2001).
\bibitem{Born}
V. Born, C. Jooss, H.C. Freyhardt, Physica C {\bf 382}, 224 (2002).
\bibitem{Murakami}
H. Murakami, H. Asaoka, K. Sakai, T. Ito, M. Tonouchi, Appl. Surf. Sci. {\bf 175-176}, 306 (2001).
\bibitem{Edwards1}
H.L. Edwards, J.T. Markert, A.L. de Lozanne, Phys. Rev. Lett. {\bf 69}, 2967 (1992).
\bibitem{Edwards2}
H.L. Edwards, D.J. Derro, A.L. Barr, J.T. Markert, 
A.L. de Lozanne, Phys. Rev. Lett. {\bf 75}, 1387 (1995).
\bibitem{Hewitt}
K.C. Hewitt, J.C. Irwin, Phys. Rev. B {\bf 66}, 054516 (2002).
\bibitem{Hoffmann}
A. Hoffmann, P. Lemmens, L. Winkeler, G. Guntherodt, J. Low Temp. Phys. {\bf 99}, 201 (1995).
\bibitem{Ponomarev}
Y.G. Ponomarev, N.Z. Timergaleev, A.O. Zabezhaylov, K.K. Uk, M.A. Lorenz, G. Muller, H. Piel, H. Schmidt, C. Janowitz, A. Krapf, R Manzke, Conference Series-Institute of Physics, {\bf 2}, 167 (2000).
\bibitem{Oki}
T. Oki, N. Tsuda, D. Shimada, Physica C {\bf 353}, 213 (2001).
\bibitem{Krasnov}
V.M. Krasnov, A. Yurgens, D. Winkler, P. Delsing, T. Claeson, Phys. Rev. Lett. {\bf 84}, 5860 (2000).
\bibitem{Gupta}
A.K. Gupta, K.-W. Ng, Phys. Rev. B {\bf 58},  R8901 (1998).
\bibitem{Kanigel}
A. Kanigel, U. Chatterjee, M. Randeria, M.R. Norman, S. Souma, 
M. Shi, Z.Z. Li, H. Raffy, J.C. Campuzano, Phys. Rev. Lett. {\bf 99}, 157001 (2007).
\bibitem{Campuzano}
J.C. Campuzano, H. Ding, M.R. Norman, H.M. Fretwell, M. Randeria, A. Kaminski, J. Mesot, T. Takeuchi, T. Sato, T. Yokoya, T. Takahashi, T. Mochiku, K. Kadowaki, P. Guptasarma, D.G. Hinks, Z. Konstantinovic, Z.Z. Li, H. Raffy, Phys. Rev. Lett. {\bf 83}, 3709 (1999).
\bibitem{Tanaka}
K. Tanaka, W.S. Lee, D.H. Lu, A. Fujimori, T. Fujii, Risdiana, I. Terasaki, D.J. Scalapino, T.P. Devereaux, Z. Hussain, Z.-X. Shen, 
Science {\bf 314}, 1910 (2006).
\bibitem{Nakaono}
T. Nakano, N. Momono, M. Oda, M. Ido, J. Phys. Soc. Jpn. {\bf 67}, 2622 (1998).
\bibitem{Oda}
M. Oda, K. Hoya, R. Kubota, C. Manabe, N. Momono, T. Nakano, M. Ido, Physica C {\bf 281}, 135 (1997).
\bibitem{McElroy}
K. McElroy, D.-H. Lee, J.E. Hoffmann, K.M. Lang, J. Lee, E.W. Hudson, H. Eisaki, S. Uchida, J.C. Davis, Phys. Rev. Lett. {\bf 94}, 197005 (2005).
\bibitem{Matsuda}
A. Matsuda, T. Fujii, T. Watanabe, Physica C {\bf 388-389}, 207 (2003).
\bibitem{Hoffman}
J.E. Hoffman, E.W. Hudson, K.M. Lang, V. Madhavan, H. Eisaki, S. Uchida, J.C. Davis, Science {\bf 295}, 466 (2002).
\bibitem{Howald}
C. Howald, P. Fournier, A. Kapitulnik, Phys. Rev. B {\bf 64}, 100504(R) (2001). 
\bibitem{Murakami1}
H. Murakami, R. Aoki, J. Phys. Soc. Jpn. {\bf 64}, 1287 (1995).
\bibitem{Almasan}
C. Almasan, M.B. Maple, in: Chemistry of High-Temperature Superconductors, edited by C.N.R. Rao (World Scientific, Singapore), 1991.
\bibitem{Biswas}
A. Biswas, P. Fournier, M.M. Qazilbash, V.N. Smolyaninova, H. Balci, R.L. Greene, 
Phys. Rev. Lett. {\bf 88},  207004 (2002).
\bibitem{Zimmers1}
A. Zimmers, R.-M. Lobo, N. Bontemps, C.C. Homes, M.C. Barr, Y. Dagan, R.L. Greene, 
Phys. Rev. B {\bf 70}, 132502 (2004).
\bibitem{Dagan}
Y. Dagan, R. Beck, R.L. Greene, Phys. Rev. Lett. {\bf 99}, 147004 (2007).
\bibitem{Homes}
C.C. Homes, R.P.S.M. Lobo, P. Fournier, A. Zimmers, R.L. Greene, Phys. Rev. B {\bf 74}, 214515 (2006).
\bibitem{Fournier}
P. Fournier, R.L. Greene, Phys. Rev. B {\bf 68}, 094507 (2003).  
\bibitem{Nishiyama}
N. Nishiyama, G. Kinoda, S. Shibata, T. Hasegawa, N. Koshizuka, M. Murakami, J. Supercond. {\bf 15}, 351 (2002).
\bibitem{Kugler3}
M. Kugler, G. Levy de Castro, E. Giannini, A. Piriou, A.A. Manuel, C. Hess, O. Fischer, J. Phys. Chem. Solids
{\bf 67}, 353 (2006).
\bibitem{Masui}
T. Masui, M. Limonov, H. Uchiyama, S. Lee, S. Tajima, A. Yamanaka, Phys. Rev. B {\bf 68}, 060506(R) (2003).
\end{thebibliography}
\end{document}